%% file: main.tex
\documentclass[11pt,letterpaper]{article}

\usepackage[T1]{fontenc}
\usepackage{lmodern}
\usepackage[margin=1in]{geometry}
\usepackage{microtype}
\usepackage{amsmath,amssymb,amsthm,bm}
\usepackage{graphicx}
\usepackage{array,booktabs,tabularx}
\usepackage{enumitem}
\usepackage{float}
\usepackage{xcolor}
\usepackage{authblk}
\usepackage[round,authoryear]{natbib}
\usepackage[hidelinks]{hyperref}

\graphicspath{{figures/}}
\setlist[enumerate,1]{label=(\alph*), ref=(\alph*), leftmargin=*}
\allowdisplaybreaks[1]

\newtheorem{thm}{Theorem}

\newtheorem{condition}{Condition}
\floatstyle{ruled}
\newfloat{algorithm}{tbp}{loa}
\floatname{algorithm}{Algorithm}

\input{self-define.tex}

\title{Harnessing Collaborative Networks\\ for Nonparametric Quantile Learning}
\author[1]{Jinyuan Chang}
\author[2]{Yue Wang}
\author[3]{Heng Lian}
\author[4]{Wen-Xin Zhou}
\affil[1]{Joint Laboratory of Data Science and Business Intelligence, Institute of Statistical
Interdisciplinary Research, Southwestern University of Finance and Economics, Chengdu, China}
\affil[2]{Faculty of Business for Science and Technology, School of Management,
University of Science and Technology of China, Hefei, China}
\affil[3]{Department of Mathematics, City University of Hong Kong, Hong Kong, China}
\affil[4]{Department of Information and Decision Sciences, University of Illinois Chicago, Chicago, USA}
\date{}

\hypersetup{
  pdftitle={Harnessing Collaborative Networks for Nonparametric Quantile Learning},
  pdfauthor={Jinyuan Chang, Yue Wang, Heng Lian, and Wen-Xin Zhou}
}

\begin{document}
\maketitle

\begin{abstract}
Data are increasingly distributed across networked agents that must learn collaboratively without exchanging raw observations or relying on central coordination. Existing decentralized methods largely focus on finite-dimensional
models with smooth and strongly convex objectives, leaving nonparametric distributional learning over networks underexplored. We study decentralized nonparametric quantile regression in reproducing kernel Hilbert spaces. The standard kernel estimator is poorly suited to this setting: the check loss is nonsmooth, while its representer theorem formulation requires a global, data-dependent basis whose dimension grows with the total sample size. We reformulate the estimator using convolution smoothing and shared random features, yielding a differentiable objective with a common, data-independent parameterization of substantially lower dimension. We develop a gradient-adjusted alternating direction method of multipliers algorithm that replaces
costly local minimization with network-corrected gradient updates, together with a communication-enhanced variant that selectively transmits updates. We establish nonasymptotic error bounds for the smoothed full-kernel and random-feature estimators, showing that the latter attains the full-kernel rate with suitable bandwidth and feature dimension. We also show both algorithms converge linearly to the random-feature consensus estimator. Simulations and a real data application demonstrate competitive predictive accuracy with substantial computational and communication savings.
\end{abstract}


\section{Introduction}
\label{sec:intro}

Modern datasets are increasingly collected and stored across geographically or organizationally distributed agents rather than in a single repository  \citep{agarwal2011distributed,zhang2015divide,jordan2019communication,fan2023communication,liu2025communication}. When pooling raw observations is restricted, distributed methods allow the agents to learn collaboratively by exchanging intermediate information. A common distributed paradigm relies on a central server to aggregate this information \citep{wang2019distributed,chen2025distributed}, but the server may become a communication bottleneck or a single point of failure. We instead consider a fully decentralized architecture in which each agent computes using its local data and communicates only with its immediate  neighbours over a prescribed network \citep{liu2022fast,wu2023network,wang2026byzantine}. Our interest is in quantile regression (QR), which characterizes different parts of a conditional distribution and is particularly useful for asymmetric, heterogeneous, and tail-sensitive responses
\citep{koenker2005quantile,li2007quantile,wang2012estimation,he2013quantile,chen2019quantile}.
Allowing the conditional quantile function to vary nonlinearly with the predictors is important in applications such as environmental monitoring and financial risk assessment, but introduces substantial statistical and computational challenges over a decentralized network.

Reproducing kernel Hilbert spaces (RKHSs) provide a flexible framework for nonparametric QR through regularized estimation \citep{li2007quantile}. The resulting kernel methods are conceptually simple and computationally tractable, and have performed competitively against neural networks across a variety of tasks. For a pooled sample of size $N$, the representer theorem expresses the standard kernel QR estimator as $f(\cdot)=\sum_{i=1}^{N}\alpha_iK(\vx_i,\cdot)$, with basis functions indexed by the total observations \citep{scholkop2001kernel}. This formulation is poorly suited to decentralized learning for three reasons. First, the check loss is nonsmooth, preventing the direct use of many decentralized algorithms whose linear-convergence theory requires differentiable local objectives with Lipschitz-continuous gradients. Although the quadratic RKHS penalty supplies strong convexity, it does not remove this nonsmoothness. Second, consensus over the coefficient vector requires all agents to use the same basis functions. Under the representer form, these basis functions are constructed from data distributed across the network and therefore are not jointly available without sharing raw data. Third, the dimension of the consensus variable grows with $N$, producing substantial computation, storage, and communication costs. The central task is therefore not simply to optimize the standard kernel estimator over a network, but to construct an alternative statistical formulation that is compatible with fast decentralized computation while retaining the accuracy of full-kernel QR.

To address this, we reformulate kernel QR so that its objective has the geometry required for fast decentralized optimization and its decision variable admits a common, data-independent finite-dimensional representation across agents. Convolution smoothing \citep{fernandes2021smoothing,tan2022high,he2023smoothed,zhou2024sparse} provides a differentiable objective with a Lipschitz-continuous gradient, while shared random features (RFs) map the local estimators into a common parameter space whose dimension can be substantially smaller than the pooled sample size \citep{rahimi2007random,rudi2017generalization,dai2023kernel}.
The key question is whether these two modifications can be jointly calibrated without changing the statistical rate of the full-kernel estimator. The bandwidth $h$ controls both approximation bias and objective smoothness; the regularization parameter $\lambda$ controls statistical complexity and supplies curvature; and the number of random features $L$ controls representation error as well as the dimension communicated over the network. Their effects must be analysed jointly rather than treated as independent computational choices.

The proposed framework addresses a gap between the statistical theory of kernel QR and decentralized kernel optimization. Shared RF representations have been used in decentralized kernel learning, while linearized and communication-censored ADMM methods provide efficient first-order updates for smooth consensus problems
\citep{richards2020decentralised,xu2021coke,ling2015dlm,liu2019communication}. These methods, however, do not provide a statistically justified smooth objective for kernel QR. In the statistical literature,  \cite{lian2022distributed} consider divide-and-conquer kernel QR, and \cite{wang2024communication} combine RF with a communication-efficient surrogate likelihood for distributed nonparametric QR. Both studies adopt a centralized distributed architecture rather than fully decentralized  neighbour-to-neighbour consensus, and neither analyses a convolution-smoothed full-kernel estimator nor the joint calibration of $h$, $\lambda$, and $L$. Smoothing has also been used algorithmically in kernel QR to approximate the solution of the original nonsmooth estimator
\citep{tang2025fastkqr}, while recent kernel QR-RF theory establishes capacity-dependent optimal rates for RF approximations of the original nonsmooth estimator \citep{wang2024optimal}. In contrast, our analysis supplies the statistical justification for the smooth RF objective required by fast decentralized optimization. By linking smoothing, regularization, RF approximation, and network computation, it identifies an estimator that is simultaneously compatible with  neighbour-to-neighbour optimization and rate-equivalent to its full-kernel counterpart.

We illustrate the proposed methodology using the Beijing Multi-Site Air-Quality dataset, which contains pollutant concentrations and meteorological measurements collected at 12 monitoring stations.
Predicting high conditional quantiles of PM$_{2.5}$ is important for assessing severe air-pollution risk, but upper-tail observations available at an individual station are necessarily limited.
We regard the stations as agents in a decentralized network: observations remain locally stored, and geographically  neighbouring stations exchange only the optimization variables required to compute a common conditional quantile estimator. Empirical results show that collaborative estimation improves upper-quantile prediction relative to local estimation and several distributed benchmarks, while the communication-enhanced procedure substantially reduces the number of transmissions.
The variable-importance analysis further reveals quantile-dependent predictive patterns: pollutant-related variables are more prominent at moderate quantile levels, whereas meteorological variables become increasingly relevant for upper-tail prediction. These findings illustrate the value of decentralized nonparametric quantile learning for air-quality monitoring and upper-tail risk assessment.

The remainder of the paper is organized as follows. Section~\ref{sec:method} introduces the proposed statistical formulation and GADMM algorithm. Section~\ref{sec:method2} presents its communication-enhanced variant. Section~\ref{sec:theory} develops the statistical and optimization theory. Section~\ref{sec:simu} reports simulation studies and a real application.
All proofs and additional technical details are provided in the separate supplementary material.

\emph{Notation.}
For any positive integer $r \ge 2$, let $[r]=\{1, \ldots, r\}$. For a vector $\vv=(v_1, \ldots, v_n)^\top$, its $\ell_d$-norm, for $d\geq 1$, is denoted by $\|\vv\|_d=(\sum_{i=1}^n|v_i|^d)^{{1}/{d}}$. For two vectors $\vu$ and $\vv$, write $\langle \vu, \vv \rangle=\vu^\top\vv$. For a positive definite matrix $\vP$, define $\|\vv\|_{\vP}=\sqrt{\vv^\top \vP\vv}$. Let $\vI_n$ denote the $n \times n$ identity matrix. For a symmetric matrix $\vA$, let $\lambda_{\min}(\vA)$ and $\lambda_{\max}(\vA)$ denote its smallest and largest eigenvalues, respectively.  For matrices $\vA_1,\ldots,\vA_m$, let $\operatorname{diag}(\vA_1,\ldots,\vA_m)$ denote the block-diagonal matrix with diagonal blocks $\vA_1,\ldots,\vA_m$.  For two positive sequences $\{a_n\}_{n\geq1}$ and $\{b_n\}_{n\geq1}$, write $a_n \lesssim b_n$, or equivalently $b_n\gtrsim a_n$, if $\limsup_{n\rightarrow\infty} a_n/b_n<\infty$. Write $a_n \asymp b_n$ if both $a_n\lesssim b_n$ and $b_n\gtrsim a_n$, and write $a_n \ll b_n$ if $\limsup_{n\rightarrow\infty} a_n/b_n =0$. For a set $S$, let $|S|$ denote its cardinality.  Let $\calX \subseteq \mathbb{R}^p$ ($p\geq 1$) denote the compact covariate space and $\mathbb P_X$ the marginal distribution of $\vx$ on $\calX$. We assume that $\mathbb P_X$ has full support on $\calX$.  For two functions $f$ and $g$ on $\calX$, define $\|f\|_2^2=\int_{\calX} f^2(\vx)\, \mathrm{d} \mathbb{P}_X(\vx)$, $\langle f, g \rangle=\int_{\calX} f(\vx)g(\vx)\, \mathrm{d}\mathbb{P}_X(\vx)$, and $\|f\|_{\infty}=\sup_{\vx \in \calX} |f(\vx)|$.
Let $L^2=\{f: \|f\|_2^2 < \infty\}$ denote the space of square-integrable functions. For an operator $\calA$, let $\|\calA\|$ and $\operatorname{Tr}(\calA)$ denote its operator norm and trace, respectively, and let $\calI$ denote the identity operator. Throughout the paper, $C$ denotes a generic positive constant whose value may change from line to line.

\section{Nonparametric Quantile Regression over Networks}
\label{sec:method}

Throughout, we consider an RKHS $\calH=\calH_k$ induced by a continuous, symmetric, and positive semidefinite kernel $k(\cdot,  \cdot): \calX \times \calX  \to \mathbb{R}$. The inner product and the norm in $\calH$ are denoted by $\langle \cdot, \cdot \rangle_{\calH}$ and $\| \cdot \|_{\calH}$, respectively, and satisfy $\langle k(\vx, \cdot), k(\vx', \cdot) \rangle_\calH  =  k(\vx, \vx')$. Equivalently, $\calH$ is the completion of the linear span of $\{ k(\vx,\cdot):\vx\in\calX\}$. Further background on RKHSs is provided in Section~D of the supplementary material.

\subsection{Problem Formulation}

Consider a decentralized network represented by a connected graph $G=(V, E)$ without self-loops, where $V=[m]$ is the set of agents and $E$ is represented as a set of directed arcs. We assume that communication is bidirectional, so that $(j,k)\in E$ if and only if $(k,j)\in E$. Agents $j$ and $k$ are called neighbours if $(j,k)\in E$. Suppose that $\{ (\vx_i,y_i) \}_{i=1}^N$, with $\vx_i\in \calX \subseteq \mathbb{R}^p$ and $y_i\in\mathbb{R}$, are independent and identically distributed (i.i.d.) observations. Agent $j\in V$ holds a local dataset indexed by $\calD_j\subseteq[N]$, where $|\calD_j|=n_j$, $\cup_{j=1}^m \calD_j=[N]$, and $\calD_j \cap \calD_{j'}=\emptyset$ for $j \neq j'$.
For simplicity, we assume that the local datasets have equal size: $\calD_j=\{(j-1)n+1, (j-1)n+2, \ldots, jn \}$, $j\in[m]$, so that the total sample size is $N=mn$. The framework can be extended to unequal local sample sizes by appropriately weighting the local empirical losses. Consider the nonparametric QR model
$$
Q_{\tau}(y_i\,|\, \vx_i)=f_{\tau}^{\star}(\vx_i)\,,
$$
where $Q_{\tau}(y_i\,|\, \vx_i)$ denotes the conditional $\tau$-th quantile of $y_i$ given $\vx_i$, $\tau \in (0,1)$, and $f_{\tau}^{\star} \in L^2$ is the unknown true regression function to be estimated. We suppress the dependence of $f_{\tau}^{\star}$ on $\tau$ for notational simplicity. Our goal is to develop a collaborative learning framework that enables the $m$ agents to estimate $f^{\star}$ jointly by solving the regularized empirical risk minimization problem
\begin{align} \label{eqn:main}
\hat{f} =\arg\min _{f \in\calH}  \bigg\{  \frac{1}{m}\sum_{j=1}^m\wh\calL_{j}(f)   +\lambda\|f\|_{\calH}^2 \bigg\} ~~\text{with}~~ \wh\calL_{ j}(f) =  \frac{1}{n}\sum_{i\in\calD_j}\rho_{\tau}(y_{i}- f(\vx_i) )\,,
\end{align}
where $\rho_{\tau}(u)=u\{\tau-I(u<0)\}$ is the check loss, and $\lambda>0$ is a regularization parameter. The local empirical loss $\wh\calL_j(\cdot)$ is computed from the data held by agent $j$ and is not directly shared with the other agents, whereas the decision function $f$ is common across the network. Because the network is connected, \eqref{eqn:main} is equivalent to the consensus problem
\begin{align}\label{eqn:model}
&  \min_{ f_1,\ldots,f_m \in\calH}  \frac{1}{m} \sum_{j=1}^m  \big\{\wh\calL_{j}(f_j)   +\lambda\|f_j\|_{\calH}^2 \big\}   \\
& ~~~~ {\rm s.t.}  ~~~~ f_j=f_k~~\hbox{for all $(j, k) \in E$}\,,  \nn
\end{align}
where $f_j$ denotes the local copy of the global decision function maintained by agent $j$. By the representer theorem, any solution to \eqref{eqn:main} admits the finite-dimensional, data-dependent representation $f(\vx) = \sum_{i=1}^N \alpha_i k(\vx, \vx_i) = \valpha^\top  \vk(\vx)$, where $\valpha=(\alpha_1, \ldots, \alpha_N)^\top$ and $\vk(\vx) = (k(\vx, \vx_1),\ldots, k(\vx, \vx_N) )^\top$. By the reproducing property, $\langle k(\cdot, \vx), k(\cdot, \vx^{\prime})\rangle_{\calH}= k(\vx, \vx^{\prime})$, we have $\|f\|_{\calH}^2=\valpha^\top \mathbf{K} \valpha$, where $\mathbf{K}=( k(\vx_i, \vx_j)) _{i, j \in [N]}$ is the kernel matrix. Consequently, \eqref{eqn:model} can be rewritten as the finite-dimensional
consensus problem
\begin{align}\label{eqn:model2}
&\min _{\boldsymbol{\alpha}_1,\ldots,\boldsymbol{\alpha}_m \in \mathbb{R}^N}   \frac{1}{m} \sum_{j=1}^m \bigg\{  \frac{1}{n} \sum_{i\in\calD_j}  \rho_{\tau}(y_i - \valpha_j^\top \vk(\vx_i))+  \lambda  \valpha_j^\top \mathbf{K} \valpha_j  \bigg\}  \\
&~~~~~ {\rm s.t.}  ~~~~~  \valpha_j=  \valpha_k~~\hbox{for all $(j, k) \in E$}\,,  \nn
\end{align}
where $\valpha_j$ denotes the local copy of the global
coefficient vector maintained by agent $j$. Compared with decentralized learning for finite-dimensional parametric models, the nonparametric QR formulation introduces several
additional challenges:

\begin{enumerate}
\item  The check loss is nonsmooth, which prevents the direct use of decentralized algorithms designed for smooth local objectives. Even though the quadratic regularization makes the overall objective strongly convex, decentralized
subgradient-type methods for the resulting nonsmooth problem
generally achieve only sublinear convergence.

\item The data-dependent representation of the decision function complicates consensus optimization. In particular, solving \eqref{eqn:model2} requires each agent to construct the kernel vector $\vk(\vx_i)\in\mathbb{R}^N$ and the kernel matrix $\mathbf K\in\mathbb{R}^{N\times N}$. This would require access to observations $\{(\vx_i,y_i):i\in\calD_k\}$, $k\in V\setminus\{j\}$, held by other agents, and is therefore incompatible with the
requirement that raw data remain local.

\item The dimension of each coefficient vector $\valpha_j$ equals the total sample size $N$. Consequently, the computational, storage, and communication costs can become prohibitive in large-scale applications.
\end{enumerate}

To address the first challenge, Section~\ref{sec:smooth} introduces convolution smoothing to construct a differentiable
approximation to the check loss, thus enabling efficient
gradient-based optimization. The strong convexity of the
overall objective is supplied separately by the quadratic
regularization. To address the remaining two challenges,
Section~\ref{sec:rf} introduces an RF approximation that
avoids the exchange of raw local data and improves scalability
in large-scale settings. Section~\ref{sec:GADMM} then develops
a gradient-adjusted ADMM algorithm for solving the resulting
smooth, finite-dimensional consensus problem.

\subsection{Convolution-smoothed Nonparametric Quantile Regression}
\label{sec:smooth}

To introduce the smoothing construction, write the local empirical check loss in \eqref{eqn:main} as
\[
n^{-1}\sum_{i \in \calD_j}\rho_{\tau}(y_i-f(\vx_i)) = \int_{-\infty}^{\infty} \rho_{\tau}(u)\,\mathrm{d}\hat F(u),
\]
where $\hat F(\cdot)$ denotes the empirical cumulative distribution function of the ``residual'' $r_i=y_{i}-f(\vx_i)$. Owing to the discontinuous nature of $\hat F(\cdot)$, the empirical check loss exhibits the same nonsmoothness  level as $\rho_\tau(\cdot)$, which motivates the use of a smooth surrogate for $\hat F(u)$. Let $K(\cdot)$ be a symmetric and non-negative kernel function satisfying $\int_{-\infty}^{\infty} K(u)\, \mathrm{d} u =1$. Given a bandwidth parameter $h>0$, we estimate the distribution function by integrating the kernel density estimator, defined as $\hat F_h(u)= \int_{-\infty}^u n^{-1}\sum_{i \in \calD_j} K_h(t - r_i) \, \mathrm{d} t$ with $K_h(u): =h^{-1}K({u}/{h})$. Substituting $\hat F_h(\cdot)$ for $\hat F(\cdot)$ yields
$$
\frac{1}{n}\sum_{i \in \calD_j} \rho_{\tau} (y_{i}-f(\vx_i) )  \approx  \int_{-\infty}^{\infty} \rho_{\tau}(u) \, \mathrm{d}  \hat F_h(u)  =  \frac{1}{n} \sum_{i \in \calD_j} \underbrace{\int_{-\infty}^{\infty} \rho_{\tau}(u)K_h(u -r_i) \,\mathrm{d}  u}_{:=\rho_{\tau, h}(r_i)}\,,
$$
where the smoothed check loss is defined as $\rho_{\tau, h}(z) := \int_{-\infty}^{\infty} \rho_{\tau}(u)K_h(u-z)\,\mathrm{d} u$. Several examples of $\rho_{\tau, h}(\cdot)$ under different kernel functions are provided in Section~B of the supplementary material. In contrast to the non-differentiability of  $\rho_{\tau}(\cdot)$, the smoothed loss $\rho_{\tau, h}(\cdot)$ is at least twice continuously differentiable. Consequently, \eqref{eqn:model}  can be approximated by
\begin{align} \label{eqn:model50}
&\min_{f_1,\ldots,f_m \in \calH}   \sum_{j=1}^m \bigg\{
\frac{1}{n}\sum_{i\in\calD_j}  \rho_{\tau, h}(y_i -  f_j(\vx_i))+  \lambda    \|f_j\|_{\calH}^2 \bigg\}  \\
&~~~~{\rm s.t.} ~~~~~f_j = f_k~~\hbox{for all $(j, k) \in E$}\,.  \nn
\end{align}
  Since the network is connected, the consensus constraints force all local copies $\{f_j\}_{j \in [m]}$ to be identical. 

\subsection{Random Feature Approximation}
\label{sec:rf}

We use random features (RFs) to construct a finite-dimensional approximation to the kernel. Suppose that $k$ is shift invariant, so that $k(\vx, \vx') = G(\vx - \vx')$ for some function $G$. By Bochner's theorem \citep{bochner2005harmonic}, if $G$ is continuous, positive definite, and absolutely integrable, then its Fourier transform $p_k(\cdot)$ is non-negative. When the kernel is scaled so that $G(0) = 1$,  $p_k(\cdot)$ can be interpreted as a probability density function \citep{rahimi2007random}. Thus,
$$
k(\vx, \vx')=\int p_k(\vw) e^{{\rm i} \lan \vw,  \vx-\vx'\ran} \, \mathrm{d} \vw :=\mathbb{E}_{\vw}\{e^{{\rm i} \lan \vw, \vx-\vx'\ran}\}=\mathbb{E}_{\vw} \{\phi_c(\vx, \vw) \phi_c^*(\vx', \vw ) \}\,,
$$
where ${\rm i} =\sqrt{-1}$, $\mathbb{E}_{\vw}(\cdot)$ denotes expectation with respect to $\vw$, $\phi_c(\vx, \vw):=e^{{\rm i} \lan \vw, \vx \ran}$, and $*$ denotes the complex conjugate. Because $k$ is real-valued, taking real parts yields $k(\vx, \vx') = \mathbb E_{\vw} \{\cos (\langle\vw, \vx- \vx'\rangle) \}$. To obtain a real-valued RF representation,  let $\vomega =(\vw,b)$, where $\vw\sim p_k$ and $b\sim\operatorname{Unif}[0,2\pi]$ are independent, and define $\phi(\vx,\vomega) = \sqrt{2}\cos (\langle\vw,\vx\rangle+b )$.
Since $\mathbb E_b \{\phi(\vx,\vomega)\phi(\vx',\vomega)\, |\,\vw \}
=\cos (\langle\vw, \vx - \vx'\rangle ),$ it follows that
$k(\vx, \vx')=\mathbb E_{\vomega}\{\phi(\vx,\vomega) \phi(\vx',\vomega)\}$.

The key idea of RF mapping is to draw $\{\vomega_l=(\vw_l,b_l)\}_{l=1}^L$ independently from the joint distribution of $(\vw,b)$, independently of the data, and approximate the kernel function by the Monte Carlo average
$$
\tilde{k}(\vx, \vx')=\frac{1}{L} \sum_{l=1}^L \phi(\vx, \vomega _l) \phi(\vx', \vomega _l)=\vphi_L(\vx)^\top \vphi_L(\vx')\,,
$$
where $\vphi_L(\vx):=L^{-1/2}(\phi(\vx, \vomega _1), \ldots, \phi(\vx, \vomega _L))^\top$, and $L$ is the number of RFs. Let $\wt \calH$ be the RKHS induced by the approximate kernel $\tilde{k}(\cdot, \cdot)$.
 The smoothed QR estimator in the RF space is defined as
\begin{align}
\label{eqn:rf_smooth_estimator}
\hat{f}_{h, \rm RF} =\arg\min _{f \in \wt \calH}  \bigg\{\frac{1}{N}\sum_{i=1}^N \rho_{\tau, h}(y_{i}- f(\vx_i) ) + \lambda \|f\|_{\wt \calH}^2 \bigg\} \,,
\end{align}
which is precisely the common consensus solution to
\begin{align} \label{eqn:model5}
&\min_{f_1,\ldots,f_m \in \wt\calH}   \sum_{j=1}^m \bigg\{
\frac{1}{n}\sum_{i\in\calD_j}  \rho_{\tau, h}(y_i -  f_j(\vx_i))+  \lambda    \|f_j\|_{\wt\calH}^2 \bigg\}  \nn\\
&~~~~{\rm s.t.} ~~~~~f_j = f_k~~\hbox{for all $(j, k) \in E$}\,.
\end{align}
Theorem \ref{thm:staerror} in Section \ref{sec:staerror} establishes the statistical convergence rate of $\hat{f}_{h, \rm RF}$.

 In addition, any function $f_j(\vx) = \valpha_j^\top \vk(\vx) \in \calH$ can be approximated by
$$
\tilde{f}_j(\vx)=\sum_{i=1}^N  \beta_{i} \vphi_L(\vx_{i})^\top \vphi_L(\vx)=\vbeta_j^\top \vphi_L(\vx) \in \wt \calH\,,
$$
where $\vbeta_j:=\sum_{i=1}^N \beta_{i} \vphi_L(\vx_{i})$ is the new decision vector in the RF space and serves as a local estimate of the global parameter $\vbeta$ at agent $j$.  Moreover, the associated norm in $\wt\calH$ satisfies

$$
\|\tilde{f}_j\|_{\wt\calH}^2 = \min_{\vbeta_j\in\mathbb R^L:
\, \tilde{f}_j(\cdot)=\vbeta_j\trans \vphi_L(\cdot)} \|\vbeta_j\|_2^2\,,
$$
where the minimum is taken over all coefficient vectors representing
$\tilde{f}_j$, since such a representation may not be unique due to possible linear dependence among $\{\phi(\cdot,\vomega_l)\}_{l=1}^L$.
Then the optimization problem \eqref{eqn:model5} reduces to estimating the parameter vectors $\{\vbeta_j\}_{j \in [m]}$:
\begin{align}\label{eqn:model4}
&\min_{\vbeta_1,\ldots, \vbeta_m \in \mathbb{R}^L}  \sum_{j=1}^m\bigg\{ \frac{1}{n} \sum_{i\in\calD_j}  \rho_{\tau, h}(y_i - \vbeta_j^\top \vphi_L(\vx_i))+ \lambda \|\vbeta_j\|_2^2 \bigg\}  \\
&~~~~~{\rm s.t.}~~~~~\vbeta_j = \vbeta_k~~\hbox{for all $(j, k) \in E$}\,. \nn
\end{align}
Throughout this paper, we use the Gaussian kernel $k(\vx, \vx')=\exp \{ -\|\vx-\vx'\|_2^2 /(2 \sigma^2) \}$ with $\sigma>0$. The associated
spectral density $p_k$ is the density of the multivariate
normal distribution $\mathbf N(\vzero,\sigma^{-2}\vI_p)$.  Under this construction, both $\vphi_L(\vx)$ and $\vbeta_j$ have fixed dimension $L$.  We assume that all agents share the same set of RFs $\{\vomega _l\}_{l=1}^L$. During communication among agents in the decentralized network, only the parameter vector $\vbeta_j$ needs to be transmitted.
The use of RF mapping is crucial, as it yields a data-independent optimization parameter $\vbeta$ shared across all agents, with dimension $L$ much smaller than the total sample size $N$. Moreover, the non-injective mapping $\vx \mapsto \vphi_L(\vx)$ provides an additional level of privacy protection.

\subsection{Implementation via Gradient-adjusted ADMM Algorithm}
\label{sec:GADMM}

To solve \eqref{eqn:model4}, we derive a gradient-adjusted ADMM algorithm. Define the smoothed empirical objective by
$$
\wh \calL_h(\vbeta)= \sum_{j=1}^m  \wh \calL_{h, j}(\vbeta_j)  ~~\text{with}~~\wh \calL_{h, j}(\vbeta_j) = \frac{1}{n}\sum_{i\in\calD_j}\rho_{\tau, h}(y_{i}-  \vbeta_j^\top\vphi_L(\vx_i)  )+ \lambda  \|\vbeta_j\|_2^2\, .
$$
To express the consensus constraints in a form suitable for ADMM, introduce an auxiliary variable $\vz_{jk}\in\mathbb R^L$ for each $(j,k)\in E$ and impose
$$
\vbeta_j = \vz_{jk}  ~~\text{and}~~  \vbeta_k =\vz_{jk}~~\text{for all} ~~(j, k) \in E\,.
$$
Let $\vbeta=(\vbeta_1^\top, \ldots, \vbeta_m^\top)^\top \in \mathbb{R}^{mL}$ collect the local parameter vectors, and let $\vz\in\mathbb R^{|E|L}$ collect the auxiliary variables
$\{\vz_{jk}:(j,k)\in E\}$. Define $\vA=(\vA_1^\top, \vA_2^\top)^\top $ and $\vB=(-\vI_{|E|L}, -\vI_{|E|L})^\top$, where $\vA_1, \vA_2 \in \mathbb{R}^{|E|L \times mL}$. Each of $\vA_1$ and $\vA_2$ consists of $|E|\times m$ blocks of size $L \times L$.  Let $r=|E|$ and let $\vchi: [r] \rightarrow E$ be a bijection. For each $e\in[r]$ such that $\chi(e)=(j,k)$, the $(e,j)$-th block of $\vA_1$ and the $(e,k)$-th block of $\vA_2$ are equal to $\vI_L$; all other blocks are zero. Section~C.3 of the supplementary material provides an illustrative five-agent example with $L=2$. With this notation, \eqref{eqn:model4} can be written as
\begin{align}\label{eqn:admm}
\min_{\vbeta \in \mathbb{R}^{mL}, ~\vz \in \mathbb{R}^{|E|L} }  \wh \calL_h(\vbeta)
~~~~~~{\rm s.t.}~~~\vA\vbeta + \vB\vz=\vzero\,.
\end{align}
For a penalty parameter $\gamma>0$, the augmented Lagrangian is
$$
L(\vbeta, \vz, \vu)= \wh \calL_h (\vbeta)+ \vu^\top( \vA\vbeta  +\vB\vz) + \frac{\gamma}{2}\|\vA\vbeta  +\vB\vz\|_2^2\,,
$$
where $\vu=(\vu_1^\top, \vu_2^\top)^\top \in \mathbb{R}^{2|E|L}$. The dual variables $\vu_1$ and $\vu_2$ correspond, respectively, to the constraints $\vA_1\vbeta -\vz=\vzero$ and $\vA_2\vbeta -\vz=\vzero$. At iteration $t$, classical ADMM alternates between
minimization over the two primal variables and a dual ascent step:
\begin{align*}
\vbeta^{(t+1)}=&\,\arg\min_{\vbeta \in \mathbb{R}^{mL}} L(\vbeta, \vz^{(t)}, \vu^{(t)})\,, \nn\\
\vz^{(t+1)} =&\,\arg\min_{\vz \in \mathbb{R}^{|E|L}} L(\vbeta^{(t+1)}, \vz, \vu^{(t)})\,, \nn\\
\vu^{(t+1)}=&\, \vu^{(t)} + \gamma(\vA\vbeta^{(t+1)} + \vB\vz^{(t+1)})\,.
\end{align*}

The main computational burden of classical ADMM arises from the $\vbeta$-update.  Because $\wh\calL_h$ is nonquadratic, this update generally requires an iterative inner optimization procedure. To avoid exact minimization, we replace $\wh\calL_h(\vbeta)$ with the proximal quadratic model
$$
\wh \calL_h(\vbeta^{(t)})+ \langle\nabla\wh \calL_h(\vbeta^{(t)}), \vbeta-\vbeta^{(t)} \rangle +\frac{1}{2}\|\vbeta-\vbeta^{(t)}\|_{\vP}^2\,,
$$
where $\vP \in \mathbb{R}^{mL \times mL}$ is a positive definite matrix to be specified below. The full gradient is $\nabla\wh \calL_h(\vbeta^{(t)}):= (\nabla \wh\calL_{h, 1}(\vbeta_1^{(t)})^\top, \ldots, \nabla \wh\calL_{h, m}(\vbeta_m^{(t)})^\top)^\top$. For each agent $j$, define $r_i(\vbeta_j)=y_{i}-\vbeta_j^\top\vphi_L(\vx_i)$ and $\calK_h(u):=\int_{-\infty}^{u/h} K(t)\,{\rm d}t$. The local gradient then becomes
\[
\nabla \wh \calL_{h, j}(\vbeta_j^{(t)})=n^{-1}\sum_{i \in \calD_j}\{\calK_h(-r_i(\vbeta_j^{(t)})) - \tau\}\vphi_L(\vx_i) + 2\lambda \vbeta_j^{(t)}.
\]
The first two terms in the proximal model form a first-order approximation to $\wh\calL_h$ at the current iterate, while
the quadratic term penalizes deviations from $\vbeta^{(t)}$. Substituting this model into the augmented Lagrangian gives
\begin{align}\label{eqn:lagrangian}
L(\vbeta, \vz^{(t)}, \vu^{(t)})  & \approx  \wh \calL_h(\vbeta^{(t)})+ \langle\nabla\wh \calL_h(\vbeta^{(t)}), \vbeta-\vbeta^{(t)} \rangle +\frac{1}{2}\|\vbeta-\vbeta^{(t)}\|_{\vP}^2 \nn\\   & \quad + \lan \vu^{(t)}, \vA\vbeta + \vB\vz^{(t)} \ran + \frac{\gamma}{2}\|\vA\vbeta  +\vB\vz^{(t)}\|_2^2\,.
\end{align}
We choose $\vP=c\vI_{mL} - \gamma \vA^\top\vA$ for some $c>0$ such that $\vP$ is positive definite. Under this choice, the coefficient of the quadratic term in $\vbeta$ is $\vP+\gamma\vA^\top\vA = c\vI_{mL}.$ Hence, the quadratic part of \eqref{eqn:lagrangian} reduces to $(c/2) \|\vbeta\|_2^2$, up to terms that are linear or constant in $\vbeta$. This choice therefore yields a closed-form update for $\vbeta$.

With the initialization $\vu_1^{(0)}=-\vu_2^{(0)}$ and $\vz^{(0)} = \vM_+\vbeta^{(0)}/2$, where $\vM_+:=\vA_1+\vA_2$, the auxiliary variable $\vz$ can be eliminated from the iterations. Moreover, the original dual variable can be reduced to $\vv=(\vv_1^\top, \ldots, \vv_m^\top)^\top:=\vM_{-}^\top\vu_1 \in \mathbb{R}^{mL}$ with $\vM_-:=\vA_1-\vA_2$. Let $\calN_j = \{k \in  V: (j, k) \in  E\}$ denote the set of neighbours of agent $j$. The resulting local updates are
\begin{align}
&\vbeta_j^{(t+1)}=\vbeta_j^{(t)} - \frac{1}{c}\bigg\{\nabla \wh \calL_{h, j}(\vbeta_j^{(t)}) + \underbrace{ \gamma \sum_{k \in \calN_j}(\vbeta_j^{(t)} -\vbeta_k^{(t)}) + \vv_j^{(t)}}_{\text{correction term}}\bigg\}\,, \label{eqn:prim_update}\\
&\vv_j^{(t+1)} = \vv_j^{(t)} + \gamma \sum_{k \in \calN_j}(\vbeta_j^{(t+1)} - \vbeta_k^{(t+1)})\,.  \label{eqn:dual_update}
\end{align}
Detailed derivations are provided in Section~C of the supplementary material. The update in \eqref{eqn:prim_update} resembles a gradient descent step with step size $1/c$, augmented by a network correction that incorporates the reduced dual variable $\vv_j^{(t)}$ and the discrepancies between agent $j$ and its neighbours. We therefore refer to the resulting procedure as the gradient-adjusted ADMM (GADMM) algorithm. GADMM requires only local computation and parameter exchange between neighbouring agents, as summarized in Algorithm~\ref{alg:gadmm}. Theorem~\ref{thm:linear} in
Section~\ref{sec:algerror} shows that $f_j^{(t)}(\vx)=\langle \vbeta_j^{(t)}, \vphi_L(\vx) \rangle$ converges to $\hat f_{h, \rm RF}$ at a linear rate.

\begin{algorithm}[!ht]
\caption{
Gradient-adjusted ADMM Algorithm.}
\label{alg:gadmm}

\small

		\begin{enumerate}[label=\arabic*:, leftmargin=2.8em, nosep]
			\item \textbf{Input}:
			\item \quad Local datasets $\{(\vx_i, y_i)\}_{i \in \calD_j}$ for agents $j \in[m]$;
			\item \quad Bandwidth $h$, regularization parameter $\lambda$, and penalty parameter $\gamma$;
			\item \quad Linearization parameter $c$, number of RFs $L$, number of iterations $T$, and Gaussian kernel bandwidth $\sigma$;

			\item \textbf{Initialization}:
			\item \quad Initialize $\vbeta_j^{(0)}$ and set $\vv_j^{(0)}=\vzero$ for every agent $j$;
			\item \quad Using a common random seed, generate $\vw_l\sim \mathbf N(\vzero,\sigma^{-2}\vI_p)$ and $b_l\sim\operatorname{Unif}[0,2\pi]$ independently for $l\in[L]$;

			\item \textbf{for} $t = 0, 1, 2, \ldots, T-1$ \textbf{do}
			\item \quad \textbf{Local primal update}:
			\item \qquad Each agent $j$ computes $\vbeta_j^{(t+1)}$ using   \eqref{eqn:prim_update};

			\item \quad \textbf{Primal variable exchange}:
			\item \qquad Each agent $j$ sends $\vbeta_j^{(t+1)}$ to every neighbour $k \in \calN_j$;
			\item \qquad Each agent $j$ receives $\vbeta_k^{(t+1)}$ from every neighbour $k \in \calN_j$;

			\item \quad \textbf{Local dual update}:
			\item \qquad Each agent $j$ computes $\vv_j^{(t+1)}$ using  \eqref{eqn:dual_update}.
			\item \textbf{end for}
				\item \textbf{Output}:  $f_j^{(T)}(\vx)=\langle \vbeta_j^{(T)}, \vphi_L(\vx)\rangle$ for each agent $j$.
		\end{enumerate}

\end{algorithm}

\section{An Enhanced Algorithm with Improved Communication Efficiency}
\label{sec:method2}

By combining convolution smoothing with RF approximation, GADMM transforms decentralized nonparametric QR into a smooth, finite-dimensional consensus problem with efficient local updates. However, as shown in \eqref{eqn:prim_update} and \eqref{eqn:dual_update}, GADMM requires agents to communicate at every iteration. Let $|\calN_j|$ denote the degree of
agent $j$. At iteration $t$, agent $j$ maintains $2+|\calN_j|$ vectors: its local primal variable $\vbeta_j^{(t)}$, its local dual variable $\vv_j^{(t)}$, and the neighbouring primal variables $\{ \vbeta_k^{(t)} \}_{k \in \calN_j}$. Although $\vv_j^{(t)}$ remains local, the updated primal variable $\vbeta_j^{(t+1)}$ must be transmitted to every
neighbour at each iteration. Repeated transmission of these $L$-dimensional vectors can result in substantial communication costs, particularly in large-scale networks.

\subsection{Communication-enhanced GADMM}
\label{sec:ce-gadmm}

To reduce communication costs, we incorporate an adaptive communication-filtering mechanism into GADMM, resulting in the communication-enhanced GADMM (CE-GADMM) algorithm. The key observation is that, when a newly computed parameter is close to the most recently transmitted value, broadcasting the update conveys little additional information and may be unnecessary. Specifically, each agent $j$ maintains a state variable $\vtheta_j^{(t)}$ that records its most recently transmitted primal variable up to iteration $t$. After computing $\vbeta_j^{(t+1)}$, agent $j$ evaluates the discrepancy $
\vr_j^{(t)}:=\vtheta_j^{(t)}-\vbeta_j^{(t+1)}$ and applies the filtering function
\begin{align}\label{eqn:censorFun}
 C_j(t, \vr_j^{(t)})=\|\vr_j^{(t)}\|_2-c_j(t)\,,
\end{align}
where  $\{c_j(t)\}$ is a nonnegative and nonincreasing threshold sequence. One convenient choice is $c_j(t)=v\mu^{t+1}$, where $v>0$ and $\mu\in(0,1)$. If $C_j(t,\vr_j^{(t)})<0$, the discrepancy is below the threshold, and the update is not transmitted. Otherwise, agent $j$ broadcasts the updated primal variable to its neighbours.

Under CE-GADMM, agent $j$ maintains $3+|\calN_j|$ vectors at each iteration $t$: its primal variable $\vbeta_j^{(t)}$, dual variable $\vv_j^{(t)}$, most recently transmitted state $\vtheta_j^{(t)}$, and the neighbouring states $\{\vtheta_k^{(t)}:k\in\calN_j\}$. Here, $\vtheta_k^{(t)}$ is the most recently received primal variable from neighbour $k$, rather than its current primal variable $\vbeta_k^{(t)}$. Thus, local computation is
performed at every iteration, whereas communication occurs only when the transmission criterion is triggered. Initialize $\vtheta_j^{(0)}=\vbeta_j^{(0)}$ for every agent $j$. At iteration $t+1$, if $C_j(t, \vr_j^{(t)})\geq 0$, agent $j$ sets $\vtheta_j^{(t+1)}=\vbeta_j^{(t+1)}$ and broadcasts $\vbeta_j^{(t+1)}$ to its neighbours. Otherwise, it sets $\vtheta_j^{(t+1)}=\vtheta_j^{(t)}$ and does not transmit an update. Similarly, when agent $j$ receives $\vbeta_k^{(t+1)}$ from a neighbour $k$, it updates the corresponding neighbour state according to $\vtheta_k^{(t+1)}=\vbeta_k^{(t+1)}$. If no update is received from agent $k$, the stored state remains unchanged: $\vtheta_k^{(t+1)}=\vtheta_k^{(t)}$. Using these most recently transmitted states, the local updates at agent $j$ become
\begin{align}
\vbeta_j^{(t+1)}=&\,\vbeta_j^{(t)} - \frac{1}{c}\bigg\{\nabla \wh \calL_{h, j}(\vbeta_j^{(t)}) + \gamma \sum_{k \in \calN_j}(\vtheta_j^{(t)} -\vtheta_k^{(t)}) + \vv_j^{(t)}\bigg\} \,, \label{eqn:prim_update2}\\
\vv_j^{(t+1)} =&\, \vv_j^{(t)} + \gamma \sum_{k \in \calN_j}(\vtheta_j^{(t+1)} - \vtheta_k^{(t+1)}) \,.  \label{eqn:dual_update2}
\end{align}
The filtering step is performed between the primal and dual updates in \eqref{eqn:prim_update2}-\eqref{eqn:dual_update2}. The algorithm listing in Section~C.1 of the supplementary material summarizes CE-GADMM, which reduces communication by suppressing below-threshold transmissions while retaining local computation at every iteration.

\subsection{Comparison with Competing Methods}

To the best of our knowledge, the existing literature does not directly address decentralized nonparametric QR. We therefore consider two natural benchmark algorithms:
decentralized gradient descent (DGD) and classical ADMM, adapted to the present setting. GADMM has the same order of per-iteration computational complexity as DGD but, under the
conditions established in Section~\ref{sec:algerror}, achieves linear convergence to the consensus solution. Compared with classical ADMM, GADMM avoids costly inner-loop optimization by replacing exact local minimization with a single gradient-based update. Its communication-enhanced extension, CE-GADMM, further reduces the number of transmissions through adaptive filtering. The main differences among these methods
are summarized in Table~\ref{t:compare}.

\begin{table}[!ht]
\centering
\caption{Comparison of DGD, ADMM, GADMM, and CE-GADMM.}
\label{t:compare}
\renewcommand{\arraystretch}{1.12}
\begin{tabularx}{\linewidth}{@{}l>{\centering\arraybackslash}X c >{\centering\arraybackslash}X c@{}}
\toprule
Method & Local update & Computation & Communication & Convergence \\
\midrule
DGD     & Gradient step       & Low  & Every iteration      & Sublinear \\
ADMM    & Exact minimization  & High & Every outer iteration & Linear \\
GADMM   & Adjusted gradient step & Low & Every iteration    & Linear \\
CE-GADMM & Adjusted gradient step & Low & Adaptively triggered & Linear \\
\bottomrule
\end{tabularx}
\end{table}

\subsubsection{Comparison with DGD}

Adapting the DGD algorithm of \cite{richards2020decentralised} to the present setting, agent $j$ updates its local parameter according to
\begin{align}\label{eqn:decen.gd}
\vbeta_j^{(t+1)} &=\sum_{k=1}^m w_{ jk} \big\{ \vbeta_k^{(t)}-\eta \nabla \wh \calL_{h,j} (\vbeta_j^{(t)}) \big\}  =\sum_{k=1}^m w_{ jk} \vbeta_k^{(t)}- \eta  \nabla \wh \calL_{h,j} (\vbeta_j^{(t)})\,,
\end{align}
where $\eta>0$ is the stepsize and $w_{jk}$ is the $(j,k)$-th entry of a nonnegative, doubly stochastic matrix $\vW=(w_{ jk}) \in \mathbb{R}^{m \times m}$ satisfying $\vW \veins = \veins$ and $\veins^\top \vW  = \veins^\top$. In addition, $w_{jk}=0$ whenever $j\neq k$ and $(j,k)\notin E$, so that each agent averages only its own iterate and those received from its neighbours. The update in \eqref{eqn:decen.gd} combines a consensus averaging step with a local gradient step. Standard DGD without gradient tracking generally requires a diminishing stepsize to converge exactly to the consensus solution,
resulting in a sublinear convergence rate; with a constant stepsize, it typically converges only to a neighbourhood of that solution \citep{Duchi2012distributed}. The GADMM update in \eqref{eqn:prim_update} has a similar gradient-based structure but includes the additional term $\vv_j^{(t)}$, which accumulates past consensus discrepancies. Thus, GADMM may be viewed as augmenting the local gradient step with a dual-memory correction. Its per-iteration computational cost is of the same order as that of DGD, because both methods require one local gradient evaluation and exchanges of $L$-dimensional parameter vectors. The additional correction, however, allows GADMM to converge linearly to the exact consensus solution under the conditions of Theorem~\ref{thm:linear}.

\subsubsection{Comparison with Classical ADMM}

Classical ADMM often attains a prescribed accuracy in
relatively few outer iterations, but each iteration may be
computationally expensive because it requires solving a
nontrivial local optimization problem. In the present
setting, the counterpart of \eqref{eqn:prim_update} is
$$
\vbeta^{(t+1)}=\arg\min_{\vbeta \in \mathbb{R}^{mL}} \bigg\{ \wh \calL_h (\vbeta)+( \vu^{(t)})^\top( \vA\vbeta  +\vB\vz^{(t)}) + \frac{\gamma}{2}\|\vA\vbeta +\vB\vz^{(t)}\|_2^2 \bigg\} \,.
$$
Because $\wh\calL_h$ is nonquadratic, this minimization
generally requires an iterative inner solver. Consequently, each outer ADMM iteration may involve multiple inner iterations, producing a nested optimization scheme with substantial computational cost. GADMM avoids this exact minimization by replacing
$\wh\calL_h$ with the proximal quadratic model in
\eqref{eqn:lagrangian}. The resulting updates
\eqref{eqn:prim_update} and \eqref{eqn:dual_update} require only a local gradient evaluation and elementary vector operations. This reduction in per-iteration computation may come at the cost of more communication rounds than classical ADMM. CE-GADMM mitigates this cost by suppressing updates whose discrepancies fall below the filtering threshold. The threshold sequence thus provides a flexible trade-off between communication frequency and optimization accuracy.

\section{Theoretical Analysis}
\label{sec:theory}

This section separates the statistical accuracy of the proposed estimators from the optimization error of the decentralized algorithms.  
We begin by introducing the operator notation used in the statistical analysis. Since $\calH$ is continuously embedded in $L^2$, let $\calS: \calH \rightarrow L^2$ denote the canonical embedding. Its adjoint $\calS^*: L^2 \rightarrow \calH$ is given by
$$
\calS^*(f)(\cdot)=\int_{\cal X} k(\vx, \cdot) f(\vx) \,  \mathrm{d}\mathbb{P}_X(\vx)\,.
$$
Define $\calV=\calS^*\calS: \calH \rightarrow \calH$ and  $\calT=\calS\calS^*: L^2 \rightarrow L^2$. The corresponding operators $\wt\calS, \wt\calS^*, \wt\calV$ and $\wt \calT$ are defined analogously for the RF-induced RKHS $\wt\calH$. The integral operator $\calT$ is self-adjoint, positive semidefinite, and trace class, and is therefore also
Hilbert-Schmidt and compact \citep{steinwart2008support}. By Mercer's theorem, there
exist an at most countable orthonormal system of eigenfunctions $\{e_j\}_{j\geq1}$ and corresponding positive eigenvalues $\mu_1 \ge \mu_2 \ge \ldots > 0$ such that
\begin{align}\label{eqn:mercer}
k(\vx, \vx^{\prime})=\sum_{j =1}^{\infty} \mu_j e_j(\vx) e_j(\vx^{\prime})~~\textrm{for any}~~\vx, \vx^{\prime} \in \calX \,.
\end{align}
For any $r >0$, the power operator $\calT^r: L^2 \rightarrow L^2$ is defined by
$\calT^r(f)(\cdot)=\sum_{j=1}^{\infty} \mu_j^r\left\langle f, e_j\right\rangle  e_j(\cdot).$


\subsection{Statistical Analysis of Nonparametric Smoothed QR in the RF Space}
\label{sec:staerror}

We first introduce the conditions.

\begin{condition}\label{cond:kernelbounded}
 The kernel function is uniformly bounded, that is, $\sup_{\vx \in \calX} k(\vx, \vx) \lesssim 1$.
\end{condition}

\begin{condition}[Source condition]\label{cond:source}
For some $r \in [1/2, 1]$, the true quantile function $f^{\star}$ satisfies $f^{\star}= \calT^r g^{\star}$ for some $g^{\star} \in L^2$ with $\|g^{\star}\|_2 \lesssim 1$.
\end{condition}

Condition~\ref{cond:source} characterizes the smoothness of the target function relative to the kernel. As $r$ increases, $f^{\star}$ becomes smoother and the corresponding function class becomes smaller. The range of $\calT^{1/2}$ coincides with the range of the canonical embedding $\calS$. Hence, $f^\star$ admits an RKHS representative whenever
$r\geq 1/2$. Even in least-squares regression, convergence rates generally cease to improve once $r>1$, a phenomenon known as the saturation effect \citep{lu2024saturation}. We therefore restrict attention to $r\in[1/2,1]$.

\begin{condition}[Conditional density]\label{cond:condensity}
Let $\varepsilon = y-f^{\star}(\vx)$, and let $f_{\varepsilon\,|\,\vx}$ denote the conditional density of $\varepsilon$ given $\vx$. There exist constants $M_1 >0$ and $0<\underline{f} \leq \bar f <\infty$ such that, almost surely with respect to $\vx$, $\underline{f}  \le \min_{|u|\le M_1} f_{\varepsilon\,|\,\vx}(u)  \le  \sup_{u \in \mathbb{R}} f_{\varepsilon\,|\,\vx}(u)  \le  \bar f$. In addition, $f_{\varepsilon \,|\, \vx}(\cdot)$ is continuously differentiable, and there exists $l_0>0$ such that $\sup_{u \in \mathbb{R}} |f_{\varepsilon \,|\, \vx}'(u)| \le l_0$ almost surely with respect to $\vx$.
\end{condition}

\begin{condition}[Smoothing kernel]\label{cond:kernel}
The smoothing kernel $K$ is continuous, symmetric, and nonnegative, and satisfies $\int_{-\infty}^{\infty} K(u) \, {\rm d} u =1$. Moreover, there exists $\bar\kappa<\infty$ such that $\sup_{u \in \mathbb{R}} K(u) \le \bar \kappa$ and $\kappa_2:=\int_{-\infty}^{\infty} u^2 K(u) \, \mathrm{d}u < \infty$.
\end{condition}

For any $\lambda > 0$, define the {\it effective dimension} of $\calT$ by
$$
\calN(\lambda)={\rm Tr}(\calT(\calT+\lambda \calI)^{-1})=\sum_{j=1}^{\infty}\frac{\mu_j}{\mu_j + \lambda} \,.
$$
The effective dimension plays a central role in determining the statistical convergence rate \citep{caponnetto2007optimal}. Under Condition~\ref{cond:kernelbounded}, $\calT$ is trace class; see, for example, Theorem~4.27 of \cite{steinwart2008support}. Consequently, $\calN(\lambda)$ is well defined for every $\lambda>0$ and
satisfies $\calN(\lambda)  \leq \lambda^{-1}\sum_{j = 1}^\infty \mu_j \lesssim \lambda^{-1}$. Moreover, $\calN(\lambda)$ is nonincreasing in $\lambda$, or equivalently, increases as $\lambda$ decreases. Since the asymptotic regime of interest has $\lambda\to 0$, we restrict attention to values of $\lambda$ for which $\calN(\lambda) \gtrsim 1$. Examples of $\calN(\lambda)$ for several commonly used
kernels are provided in the kernel summary table in Section~E of the supplementary material.

\begin{condition}\label{cond:rf.bounded}
The feature map $\phi(\cdot, \cdot)$ is continuous in both arguments and uniformly bounded: $\sup_{\vx, \vomega} |\phi(\vx, \vomega)| \lesssim 1$.
\end{condition}

Condition~\ref{cond:rf.bounded} is standard in the RF literature \citep{li2021rf,xu2021coke,dai2023kernel}. In particular, it implies that both the original kernel $k$ and its RF approximation $\tilde k$ are uniformly bounded.



\begin{thm}\label{thm:staerror}
Suppose that Conditions {\rm\ref{cond:kernelbounded}--\ref{cond:rf.bounded}} hold. For a given $t \ge 1$, assume that $h^2 \le \lambda^r$, $L\geq C \lambda^{-1} \calN(\lambda)^{2r-1} t$ for a sufficiently large constant $C>0$, and that $\lambda^{r-1/2}\lesssim 1$, $\lambda \geq \{ \calN(\lambda) + t \}/N$.  Then, with probability at least $1-C\calN(\lambda)e^{-t}$, jointly over the observed sample and the RF draw,
$$
\|\hat{f}_{h, \rm RF} - f^{\star}\|_2^2 \lesssim \lambda^{2r} + \frac{\calN(\lambda) + t}{N} \,.
$$
\end{thm}

Theorem \ref{thm:staerror} shows that the RF estimator attains the same statistical rate as the full-kernel estimator, provided that
$L$ satisfies the stated lower bound. Thus, the RF approximation reduces computational and communication costs without deteriorating the statistical rate. The theorem also makes explicit the trade-off among $\lambda$, $h$, $\calN(\lambda)$, and $L$. A smaller regularization parameter reduces the bias
$\lambda^{2r}$ but increases both the effective dimension and the number of RFs required to control the approximation error. At the same time, the condition $h^2\leq\lambda^r$ ensures that the smoothing error remains no larger than the
regularization bias. Several concrete consequences are given in
the corollaries in Section~E of
the supplementary material. In particular, suppose that the
kernel eigenvalues decay polynomially: $\mu_j\asymp j^{-\alpha}$ for some $\alpha>1$. Under the source condition in Condition~\ref{cond:source}, the polynomial-decay corollary there yields a sharper convergence rate than the corresponding result of \cite{xu2021coke}. This improvement results from the general
source condition and the refined control of the RF approximation error. The present analysis also addresses the additional difficulty introduced by QR. Unlike least-squares estimators, the smoothed quantile estimator does not admit a closed-form representation. The analyses of \cite{richards2020decentralised} and \cite{xu2021coke}, by contrast, focus primarily on smooth least-squares-type problems and exploit their explicit solution structure. They also require a bounded second moment of the response. Theorem~\ref{thm:staerror} does not impose such a moment condition, making it applicable to heavy-tailed response distributions that satisfy the conditional-density conditions stated above.

\subsection{Linear Convergence of GADMM and CE-GADMM over Networks}
\label{sec:algerror}

We now study the algorithmic convergence of GADMM from an optimization perspective. Unlike the preceding statistical analysis, the results in this subsection are conditional on the observed data and the generated RFs. Let $\calD = \{(\vx_i, y_i)\}_{i=1}^N$ denote the full dataset, and let $\vOmega =
\{\vomega_l\}_{l=1}^L$ denote the generated RF parameters. We condition throughout on a fixed realization of $(\calD,\vOmega)$. Consequently, for each agent $j$, the local empirical objective $\wh\calL_{h,j}(\vbeta_j) = n^{-1}\sum_{i\in\calD_j} \rho_{\tau,h}(y_i - \vbeta_j^\top\vphi_L(\vx_i)) + \lambda\|\vbeta_j\|_2^2$ is a deterministic function of $\vbeta_j$. The sample size $n$, bandwidth $h$, regularization parameter $\lambda$, and number of RFs $L$ are therefore treated as fixed.  The core question
reduces to how the choice of optimization parameters $c$, $\gamma$, and the network topology influences the algorithm's
convergence behaviour. Direct calculations yield the Hessian of the
empirical loss: $\nabla^2\wh \calL_{h, j}(\vbeta_j)=n^{-1}\sum_{i\in\calD_j} K_h(\vbeta_j^\top\vphi_L(\vx_i)-y_i)\vphi_L(\vx_i)\vphi_L(\vx_i)^\top + 2\lambda\vI_L$. This implies that the strong convexity and smoothness constants of  $\wh \calL_{h, j}(\vbeta_j)$ can be taken as $m_{j, \beta}= 2\lambda$ and $M_{j, \beta}=(\bar \kappa /h) \lambda_{\max}(n^{-1}\sum_{i\in\calD_j} \vphi_L(\vx_i)\vphi_L(\vx_i)^\top) + 2\lambda$, respectively; that is, for any $\vu\in\mathbb R^L$,  $m_{j, \beta}\|\vu\|_2^2 \le  \vu^\top \nabla^2 \wh\calL_{h,j}(\vbeta_j) \vu \le  M_{j, \beta} \|\vu\|_2^2$.
Since the Hessian of the aggregated empirical loss $\wh\calL_h(\vbeta)= \sum_{j=1}^m  \wh \calL_{h, j}(\vbeta_j)$ with $\vbeta=(\vbeta_1^\top, \ldots, \vbeta_m^\top)^\top$ is block diagonal:
$$
\nabla^2 \wh\calL_h(\vbeta) = \operatorname{diag}\big( \nabla^2 \wh\calL_{h,1}(\vbeta_1), \; \nabla^2 \wh\calL_{h,2}(\vbeta_2), \; \ldots, \; \nabla^2 \wh\calL_{h,m}(\vbeta_m) \big)\,,
$$
the overall strong convexity and smoothness constants for $\wh \calL_{h}(\vbeta)$ are  $m_{\beta}=\min_{j \in [m]}m_{j, \beta}$ and $M_{\beta}=\max_{j \in [m]}M_{j, \beta}$.

\begin{thm} \label{thm:linear}
Assume that Conditions {\rm\ref{cond:kernel}-\rm\ref{cond:rf.bounded}} hold, and that $\vu_1^0$ lies in the column space of $\vM_-$.  Let $\vP:=c\vI_{mL}-\gamma\vA^\top\vA$ and $\wt \vP:=c\vI_{mL} - ({\gamma}/{2})\vM_-^\top \vM_-$, where $c>0$ is chosen such that both $\vP$ and $\wt\vP$ are positive definite. Let $\sigma_1$ and $\sigma_2$ denote the smallest and largest eigenvalues of $\wt\vP$, respectively, and let $\sigma_3$ denote the smallest nonzero eigenvalue  of $(1/2)\vM_-^\top \vM_-$. Define
$$
\eta = \min \bigg\{\frac{\sigma_3\gamma(2m_{\beta}-\theta)}{\sigma_2\sigma_3\gamma + 2 \mu_0 M_{\beta}^2},  ~  \frac{\sigma_3\gamma(\mu_0-1)(\sigma_1-M_{\beta}^2/\theta)}{\mu_0\sigma_2^2+2(\mu_0-1)\mu_0 M_{\beta}^2}\bigg\} >0
$$
for some constants $\theta>0$ and $\mu_0 >1$. Then we have
$$
\max_{j \in [m]} \; \|f_j^{(t)} - \hat{f}_{h, \rm RF}\|_2  \lesssim  \Big(\frac{1}{1+\eta} \Big)^{t/2} \bigg(\frac{\sigma_2\|\vbeta^{(0)}-\wh\vbeta\|_2^2}{\sigma_1} + \frac{2\|\vu_1^{(0)} -\wh\vu_1\|_2^2}{\gamma \sigma_1} \bigg)^{1/2}\,,
$$
where $\wh \vbeta$ is the optimal primal solution, and $\wh\vu=(\wh\vu_1^\top, \wh\vu_2^\top)^\top$ is the optimal dual variable.
\end{thm}

Theorem \ref{thm:linear} establishes a linear convergence rate, which is determined by the network structure (through $\vM_-$), the smallest strong convexity constant, and the largest smoothness constant of the local empirical loss functions.  Here $\theta$ is an arbitrary positive constant satisfying $M_{\beta}^2/\sigma_1 < \theta < 2 m_{\beta}$, which is possible provided
$$
c > \max\big\{ \gamma\lambda_{\max}(\vA^\top\vA), (\gamma /2) \lambda_{\max}(\vM_-^\top \vM_-) + M_{\beta}^2/(4\lambda)\big\}.
$$

\begin{thm} \label{thm:enhan_linear}
Assume that Conditions {\rm\ref{cond:kernel}-\rm\ref{cond:rf.bounded}} hold,  $0 < \gamma < 2c_1c_2/(c_1+c_2)$ for some positive constants $c_1, c_2$, and that $\vu_1^{(0)}$ lies in the column space of $\vM_-$.  Let $c_j(t)=v \mu^{t+1}$ in \eqref{eqn:censorFun} with constants $v>0$ and $\mu \in  (0,1)$. Define $\vP:=c\vI_{mL}-\gamma\vA^\top\vA$ and $\wt \vP=c\vI_{mL} - ({\gamma}/{2})\vM_-^\top \vM_-$ for some $c>0$ such that both $\vP$ and $\wt\vP$ are positive definite. Let $\sigma_1$ and $\sigma_2$ denote  the  smallest and largest eigenvalues  of $\wt\vP$, respectively. Let $\sigma_3$ and $\sigma_4$ denote the smallest nonzero eigenvalue and largest eigenvalue of $(1/2)\vM_-^\top \vM_-$, respectively. Define 
\begin{align*}
\eta &= \min \bigg\{\frac{\gamma(2\sigma_3 c_1 m_{\beta}-c_1\sigma_3\theta -  c_3 M_{\beta}^2)}{c_1(\sigma_3\sigma_2\gamma + 2  c_3 M_{\beta}^2)}, \; \frac{1}{\mu^2} -1.01, \\
& \hspace{17mm} \frac{\gamma\{c_1\sigma_3\sigma_1( c_3-1)-c_1\sigma_3( c_3-1)M_{\beta}^2/\theta  - ( c_3-1) c_3 M_{\beta}^2 -  c_3 \sigma_2^2\}}{2c_1 c_3\{\sigma_2^2+( c_3-1) M_{\beta}^2\}}\bigg\} >0\,,
\end{align*}
for some constants $\theta>0$ and $c_3 >1$.  Then we have
$$
\max_{j \in [m]} \; \|f_j^{(t)} - \hat f_{h, \rm RF}\|_2  \lesssim  \Big(\frac{1}{1+\eta} \Big)^{\frac{t}{2}} \bigg\{\frac{\sigma_2 \|\vbeta^{(0)}-\wh\vbeta\|_2^2}{\sigma_1} + \frac{2\|\vu_1^{(0)}-\wh\vu_1\|_2^2}{\gamma \sigma_1 } + \frac{2r_0 mv^2/\sigma_1}{1-(1+\eta)\mu^2} \bigg\}^{\frac{1}{2}} \,,
$$
where $r_0:=\{ 2c_1 + 10c_2 + \gamma + \frac{4c_1\eta + 2\gamma}{c_1\sigma_3( c_3-1)} \gamma c_3\sigma_4  \} \sigma_4$.
\end{thm}

Theorem~\ref{thm:enhan_linear} establishes that CE-GADMM converges linearly when the filtering threshold is chosen appropriately. Relative to GADMM, the use of stale information contributes the additional term $\frac{2r_0mv^2/\sigma_1}{1-(1+\eta)\mu^2}$ to the convergence bound, which remains finite provided $(1+\eta)\mu^2 < 1$. In return, the filtering mechanism substantially reduces the frequency of communication. As demonstrated by the numerical results in Section~\ref{sec:simu}, CE-GADMM considerably lowers the overall communication cost relative to GADMM, making it well suited to applications in which network communication is a primary bottleneck.

\section{Numerical Studies}
\label{sec:simu}

In this section, we provide numerical illustrations of the proposed methods. We begin with simulation studies presented in Section~\ref{sub:simu}, followed by a real application in Section~\ref{sub:real}.

\subsection{Simulations}
\label{sub:simu}

We generate data for the $j$-th agent according to the heteroscedastic model $y=f^{\star}(\vx)+0.5 \|\vx\|_2\{\epsilon - F^{-1}_{\epsilon}(\tau)\}$, where $f^{\star}(x)=x\sin (3\pi x/2)$ for $p=1$, and $f^{\star}(\vx)=(1-\|\vx\|_2)^2$ when $p=3$. Here, $F^{-1}_{\epsilon}(\tau)$ denotes the $\tau$-th quantile of $\epsilon$, so that the conditional $\tau$-th quantile of $y$ is exactly $f^{\star}$. Unless otherwise specified, the simulation setting is as follows. The covariate $\vx$ is generated from the uniform distribution on $[0, 2]^p$, and $\epsilon$ is drawn from an exponential distribution with rate one. We set $\gamma=0.01$, $c=1$, and the smoothing parameter $h=0.1$, and consider an Erd\"{o}s-R\'{e}nyi graph with connection probability 0.5. RFs are used to approximate the Gaussian kernel with parameter $\sigma=0.2$. The filtering function is specified as $c_j(t)=v \mu^{t+1}$ with $v=100$ and $\mu=0.95$, and the quantile level is varied over $\{ 0.3 , 0.5 , 0.7\}$. The regularization parameter $\lambda$ is selected using a validation set approach. We run 1000 iterations, which we found sufficient to ensure convergence. To evaluate the performance of the proposed methods, we record the estimation error defined as the $L_2$ norm between the estimated and true functions, as well as the objective value, the computational cost measured by average running time (in minutes), and the communication cost measured by the cumulative number of transmissions. All reported results are averaged over 20 independent trials. The results for the case $p=1$ are presented in the main text, while those for $p=3$ are deferred to the supplementary material, as they exhibit similar overall patterns.

\begin{figure}[!ht]
\centering
\includegraphics[width=.32\linewidth,height=1.6in,keepaspectratio]{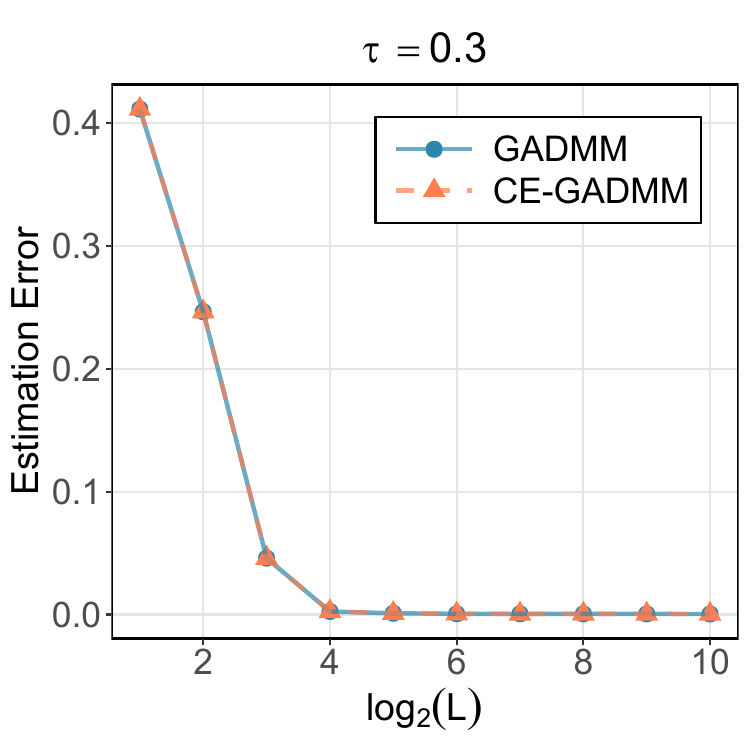}\hfill
\includegraphics[width=.32\linewidth,height=1.6in,keepaspectratio]{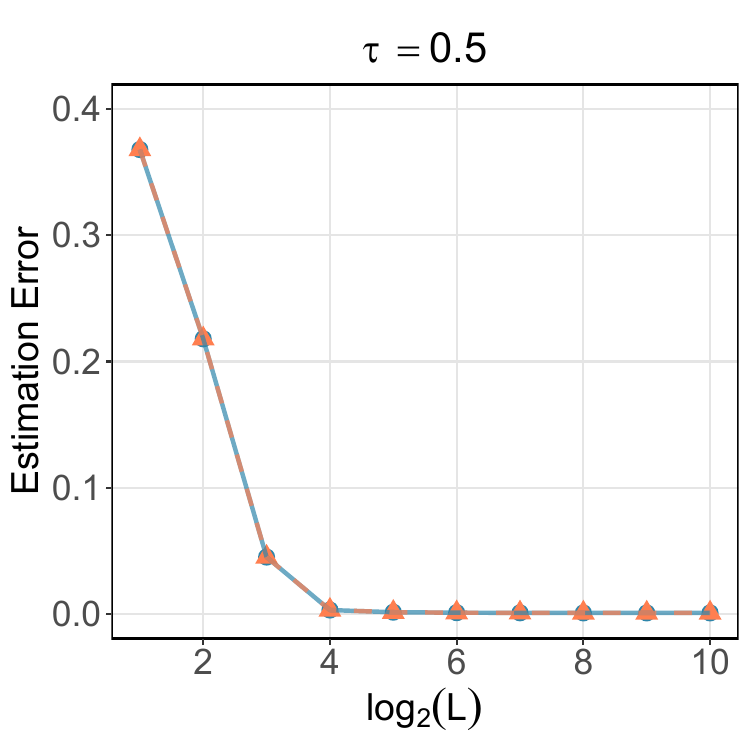}\hfill
\includegraphics[width=.32\linewidth,height=1.6in,keepaspectratio]{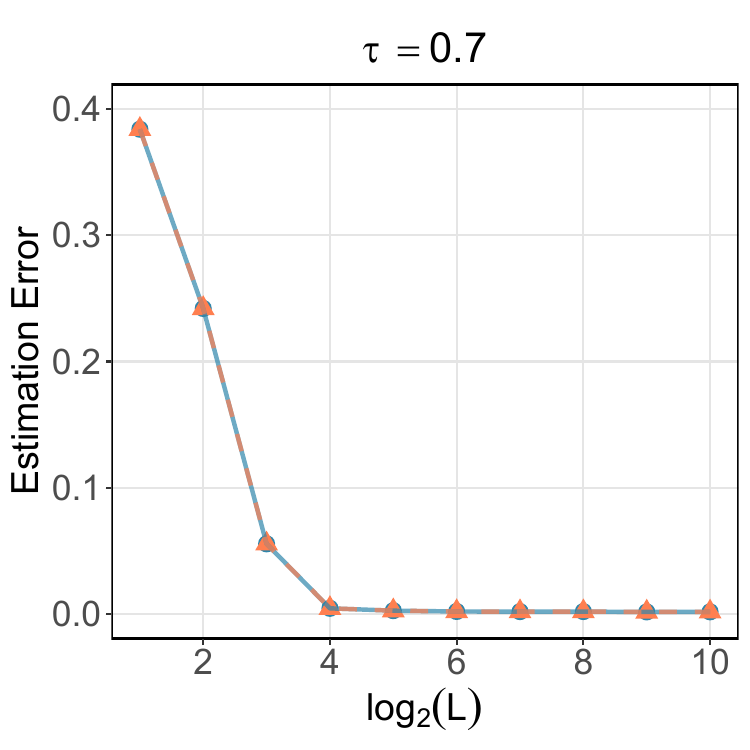}\par\medskip
\includegraphics[width=.32\linewidth,height=1.6in,keepaspectratio]{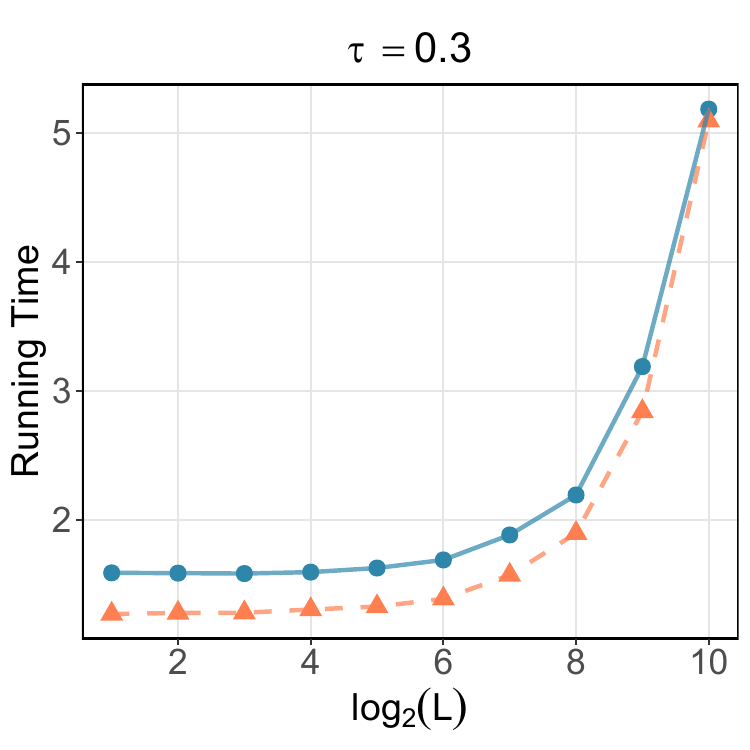}\hfill
\includegraphics[width=.32\linewidth,height=1.6in,keepaspectratio]{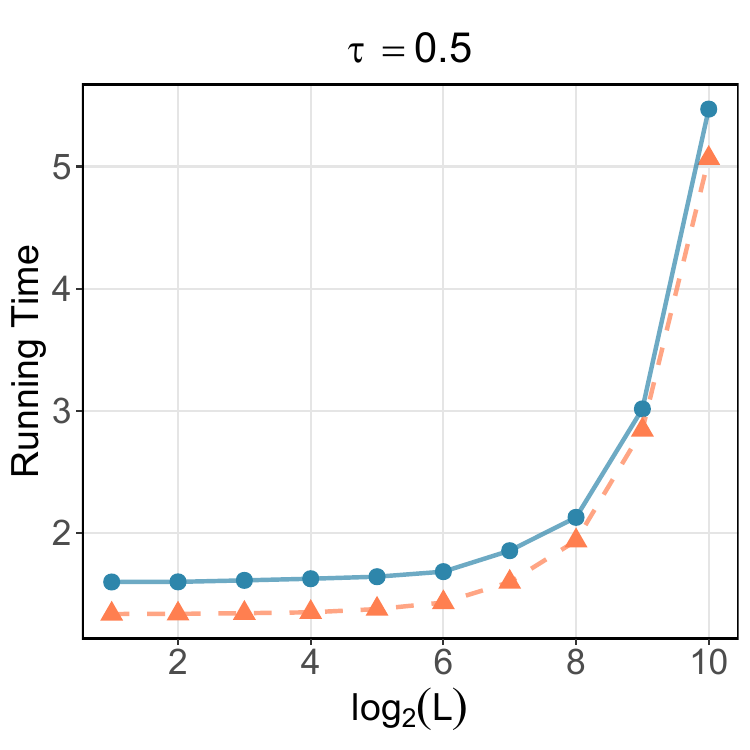}\hfill
\includegraphics[width=.32\linewidth,height=1.6in,keepaspectratio]{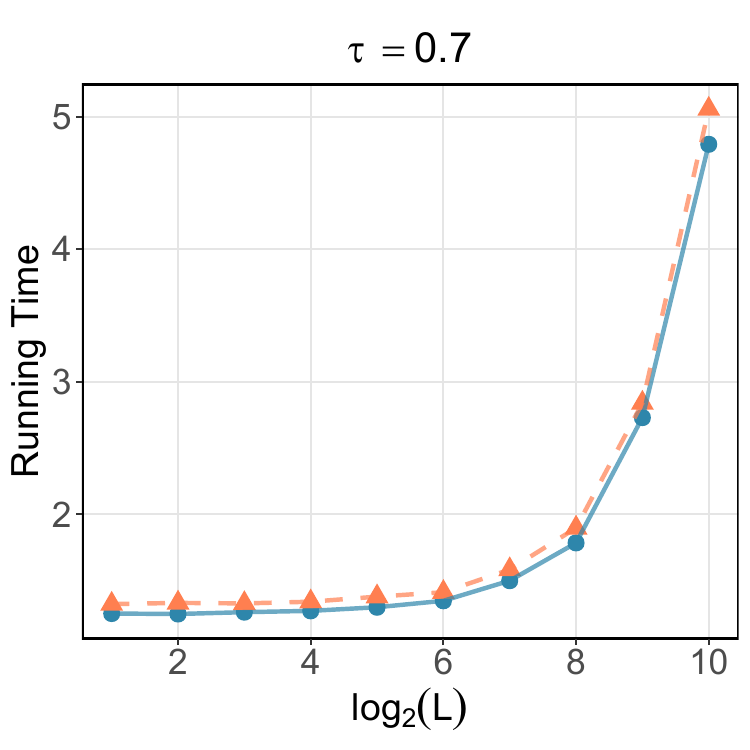}
\caption{Estimation errors and running time (in minutes) of GADMM and CE-GADMM for different numbers of random features, with $L \in \{2, 2^2, 2^3, \ldots, 2^{10}\}$. }\label{fig:RF}
\end{figure}

We set $n=1024$ and $m=10$, and vary the number of RFs over $L \in \{2, 2^2, 2^3, \ldots, 2^{10}\}$. As shown in Figure \ref{fig:RF}, the estimation error decreases as $L$ increases. Even when $L=16$ (so that $L \ll n$), the error is already visually indistinguishable from that of the standard method. Moreover, GADMM and CE-GADMM exhibit nearly identical performance, with no visible separation between their error curves. The running time comparison in the bottom panel further shows that smaller values of $L$ can substantially reduce the computational cost while maintaining high accuracy. Accordingly, we set $L=2^5$ in the subsequent simulations.

\begin{figure}[!ht]
\centering
\includegraphics[width=.32\linewidth,height=1.6in,keepaspectratio]{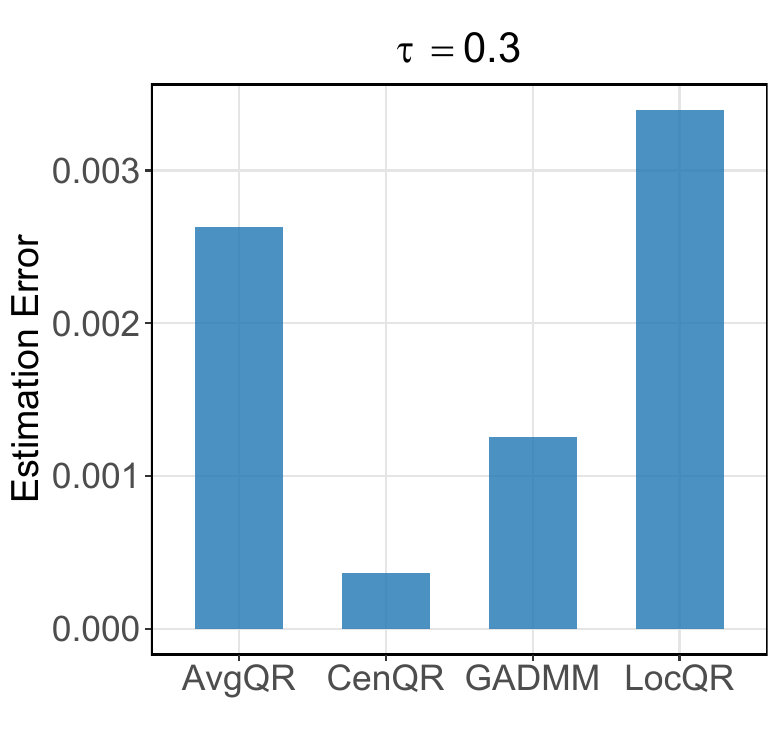}\hfill
\includegraphics[width=.32\linewidth,height=1.6in,keepaspectratio]{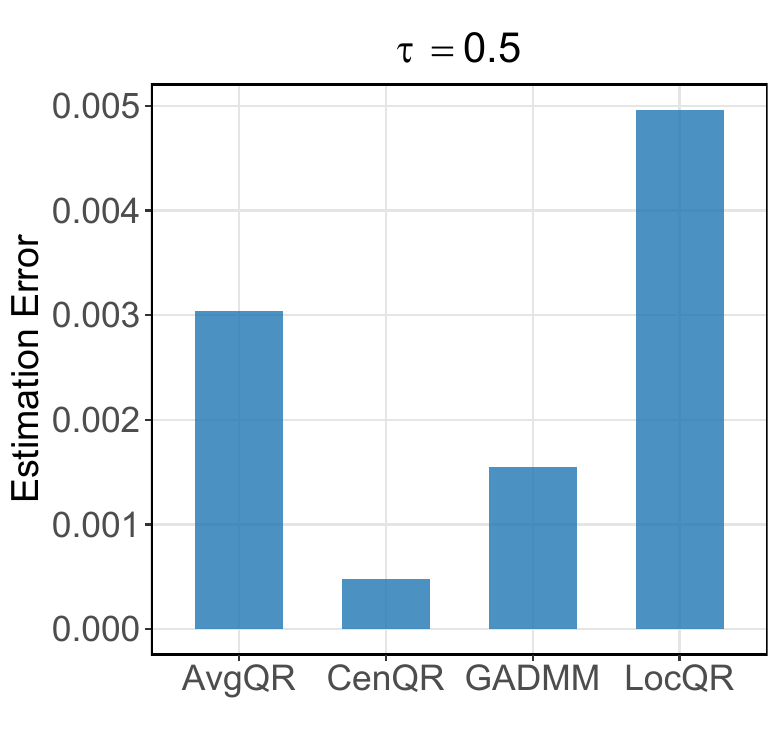}\hfill
\includegraphics[width=.32\linewidth,height=1.6in,keepaspectratio]{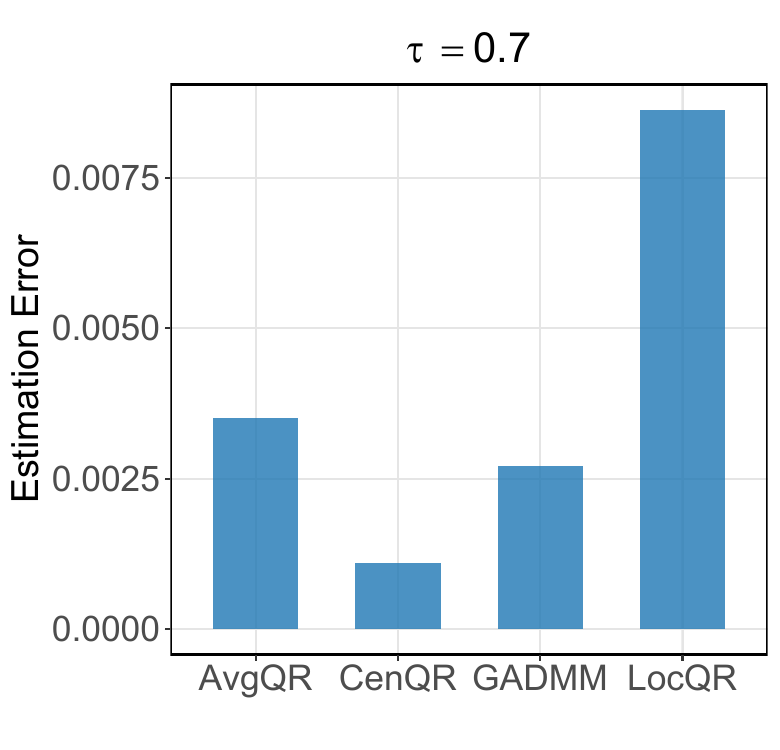}
\caption{Comparison with centralized quantile regression (CenQR), local quantile regression (LocQR), and the naive average of the local quantile estimators (AvgQR).}
\label{fig:agent}
\end{figure}

We compare the proposed methods with three procedures:
(i) CenQR: centralized QR based on aggregating all local data at a central server; (ii) LocQR: local QR carried out by a single agent using only its own $n$ samples and without communication; and (iii) AvgQR: the naive average of the local quantile estimators across all agents. Although these methods do not respect the distributed network structure, they provide useful reference points. Figure \ref{fig:agent} shows that GADMM achieves estimation accuracy closest to that of CenQR, outperforming LocQR and AvgQR. In particular, LocQR relies solely on local data, while the naive averaging method AvgQR yields only modest improvement. Although CenQR serves as an ideal benchmark by pooling all observations, it is often impractical in real applications because of the communication burden and privacy constraints discussed in Section~\ref{sec:intro}.

We next compare the proposed method with classical ADMM and decentralized gradient descent (DGD) given in \eqref{eqn:decen.gd}. For DGD, we use a stepsize of 0.5, chosen to ensure convergence while achieving a relatively fast convergence rate. We report only the results from the first 100 iterations. As shown in Table~\ref{t:compare_num}, ADMM and GADMM exhibit very similar convergence behaviour, whereas DGD converges the most slowly. However, ADMM incurs a substantially higher computation cost, roughly three times that of GADMM, because each iteration requires solving a local optimization subproblem. In terms of communication efficiency, CE-GADMM requires only 276 transmissions, compared with 1000 for GADMM. This gain arises because CE-GADMM triggers fewer transmissions during the learning process, thereby substantially improving communication efficiency.

\begin{table}[!ht]
\centering
\caption{Comparison with classical ADMM and DGD at $\tau=0.3$.}
\label{t:compare_num}
\begin{tabular}{@{}lccc@{}}
\toprule
 & Convergence & Computation cost & Communication cost \\
\midrule
\begin{tabular}[c]{@{}l@{}}
DGD \\[.8em]
ADMM \\[.8em]
GADMM \\[.8em]
CE-GADMM
\end{tabular}
& \raisebox{-.5\height}{\includegraphics[width=.26\linewidth,height=1.5in,keepaspectratio]{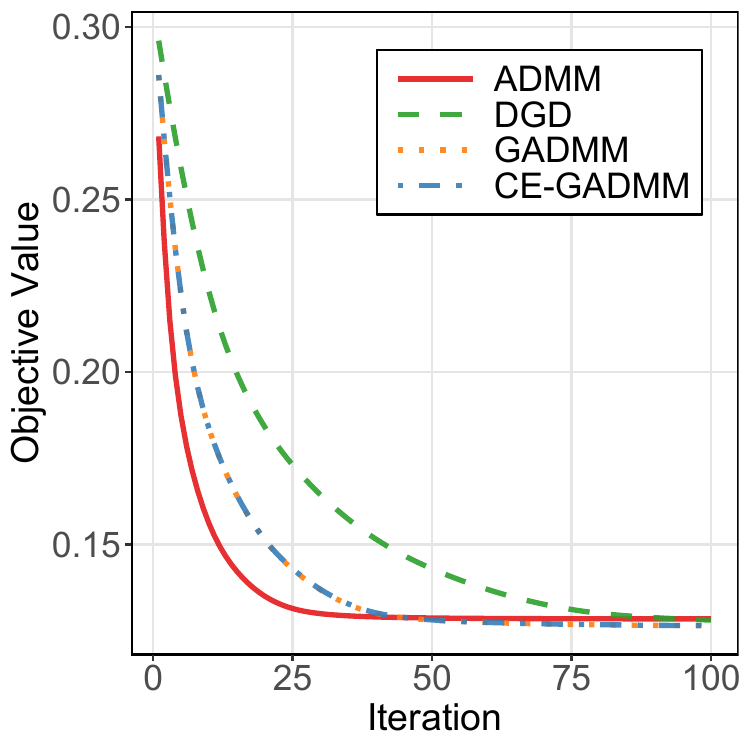}}
& \raisebox{-.5\height}{\includegraphics[width=.26\linewidth,height=1.5in,keepaspectratio]{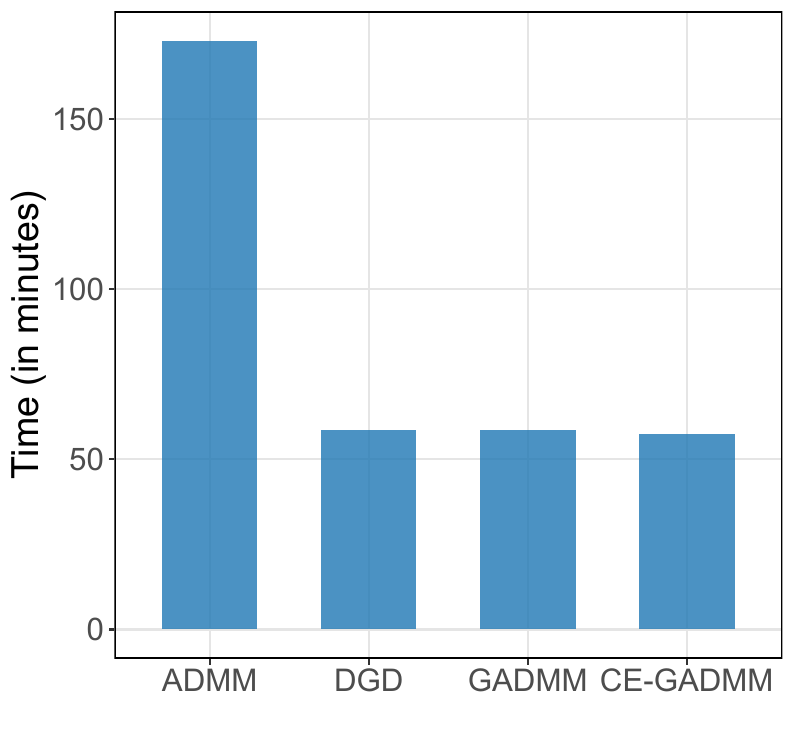}}
& \begin{tabular}[c]{@{}c@{}}
$-$ \\[.8em]
$-$ \\[.8em]
1000 \\[.8em]
276
\end{tabular} \\
\bottomrule
\end{tabular}
\end{table}

\subsection{An Application to Beijing Multi-Site Air-Quality Data}
\label{sub:real}

\begin{figure}[!ht]
	\centering
	\includegraphics[height=2.5in,width=\linewidth,keepaspectratio]{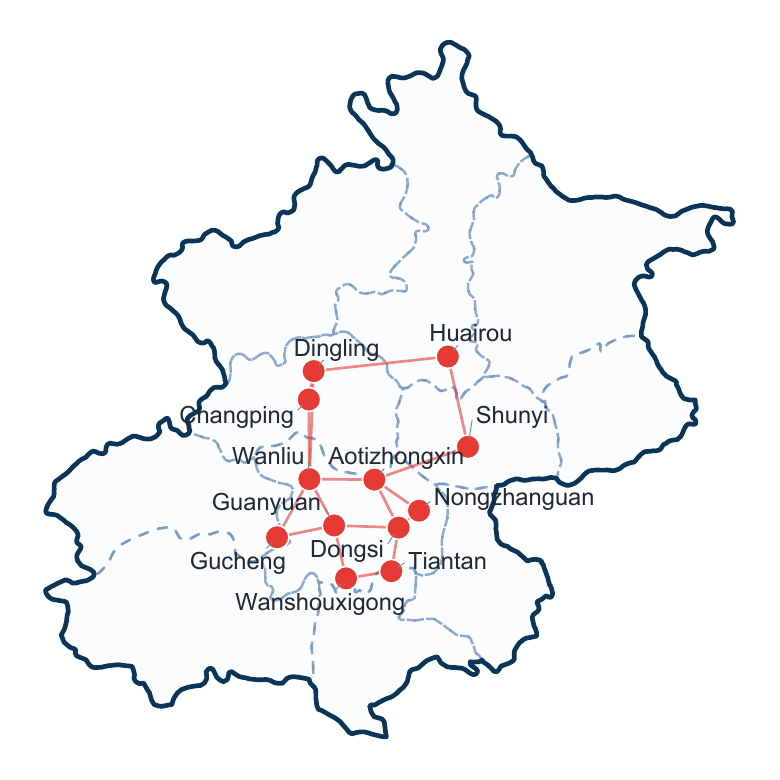}

	\caption{Network topology of the Beijing air-quality monitoring stations.}
    \label{fig:realNetwork}
\end{figure}

We apply the proposed methods to the Beijing Multi-Site Air-Quality dataset\footnote{\url{https://archive.ics.uci.edu/dataset/501/beijing+multi+site+air+quality+data}}, which
contains hourly pollutant concentrations and meteorological measurements collected at 12 monitoring stations in Beijing from March 1, 2013, to February 28, 2017.  After interpolating missing values and aggregating the hourly measurements to the daily level, we obtain $n=1,461$ observations for each station. The spatially distributed nature of the data provides a natural setting for decentralized learning.
We regard the $m=12$ monitoring stations as decentralized agents connected through a
geographical communication network. Specifically, we calculate pairwise
geographical distances based on station coordinates and connect each station with its two
geographically closest stations. To ensure bidirectional communication, the graph is converted into an undirected network by connecting two stations whenever either one belongs to the other's 2-nearest-neighbour set. The resulting network topology is displayed in Figure~\ref{fig:realNetwork}. Importantly, communication is restricted to neighbouring stations, and no raw air-quality measurements are shared.

Our response variable is the daily average concentration of PM$_{2.5}$, expressed in units of $10 \mu\mathrm{g}/\mathrm{m}^3$. PM$_{2.5}$ is a major air pollutant consisting of fine particulate matter that can penetrate deeply into the respiratory system and bloodstream, making accurate characterization of high pollution levels particularly important for environmental monitoring and public-health management \citep{yeatts2007coarse}.
Therefore, we investigate  quantiles at $\tau=0.5,0.8,0.9$ and $0.95$, where the upper quantiles correspond to increasingly severe pollution. We use eight daily aggregated variables as predictors: the concentrations of sulfur dioxide (\texttt{SO$_2$}), nitrogen dioxide (\texttt{NO$_2$}), carbon monoxide (\texttt{CO}), and ozone (\texttt{O$_3$}), together with temperature (\texttt{TEMP}), atmospheric pressure (\texttt{PRES}), dew-point temperature (\texttt{DEWP}), and wind speed (\texttt{WSPM}). All predictors are rescaled to $[0,1]$.
At each agent, the observations are randomly split into training and testing sets in a 7:3 ratio,  and this procedure is repeated over 10 independent trials. We report the average prediction error,
$
\text{PreErr}= \sum_{i=1}^{N_{\text{test}}}\rho_{\tau}(y_i-\hat{y}_i)/ N_{\text{test}}
$
on testing sets, together with the average computation time measured in minutes. Here $y_i$ denotes the observed daily average PM$_{2.5}$ concentration for the $i$th test observation, while $\hat y_i=\hat f(\vx_i)$ denotes its predicted conditional $\tau$-th quantile. Tuning parameters are chosen according to the settings in Section \ref{sub:simu}.

\begin{figure}[!ht]
\centering
\includegraphics[width=.242\linewidth,height=1.75in,keepaspectratio]{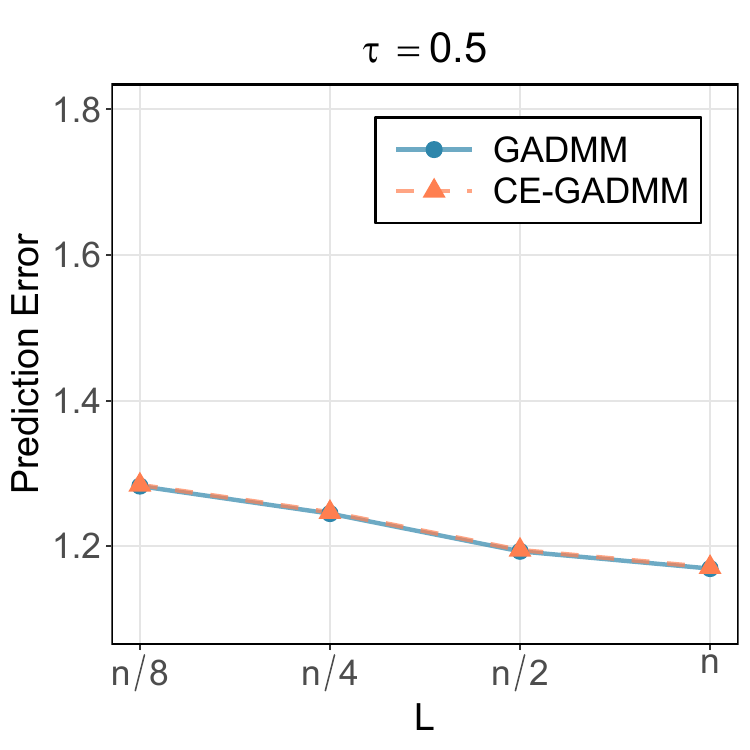}\hfill
\includegraphics[width=.242\linewidth,height=1.75in,keepaspectratio]{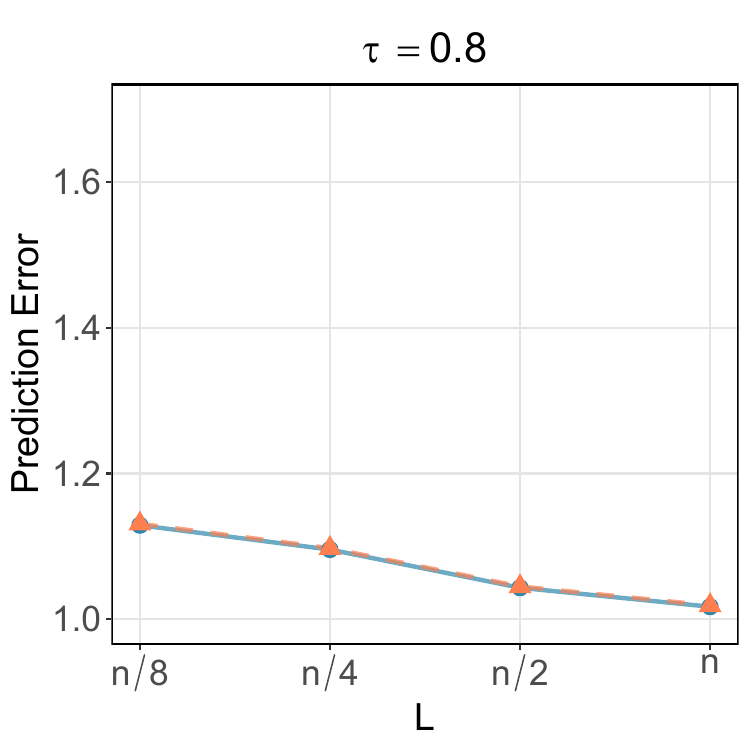}\hfill
\includegraphics[width=.242\linewidth,height=1.75in,keepaspectratio]{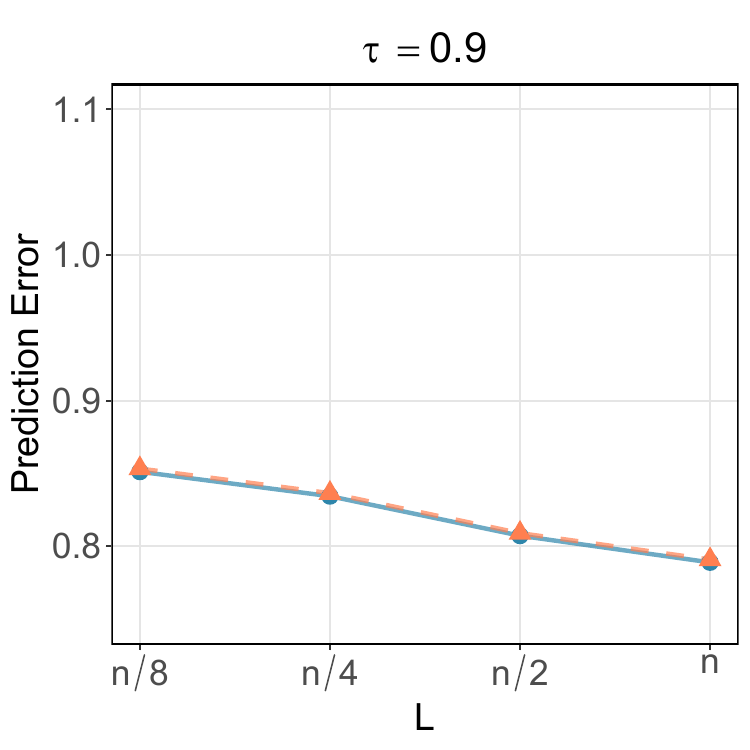}\hfill
\includegraphics[width=.242\linewidth,height=1.75in,keepaspectratio]{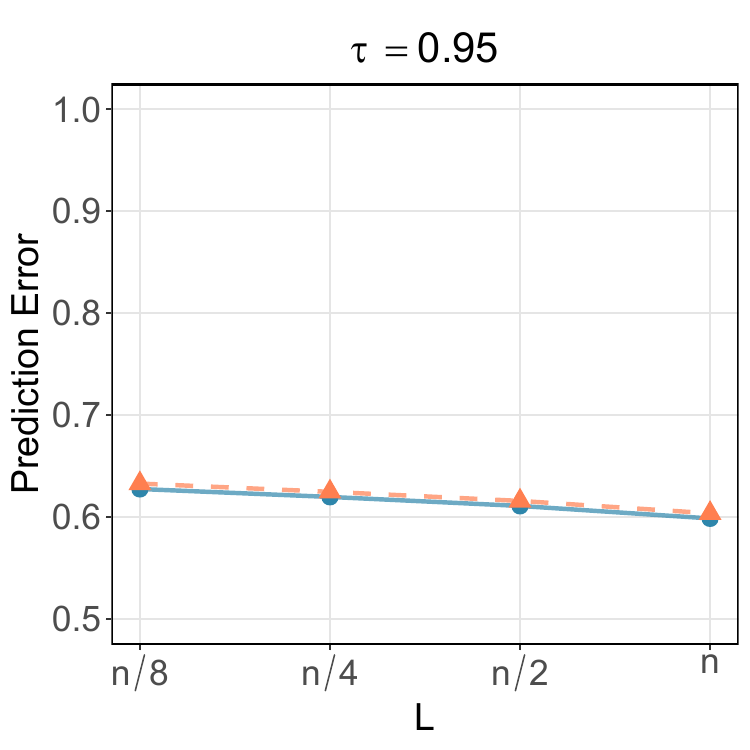}\par\medskip
\includegraphics[width=.242\linewidth,height=1.75in,keepaspectratio]{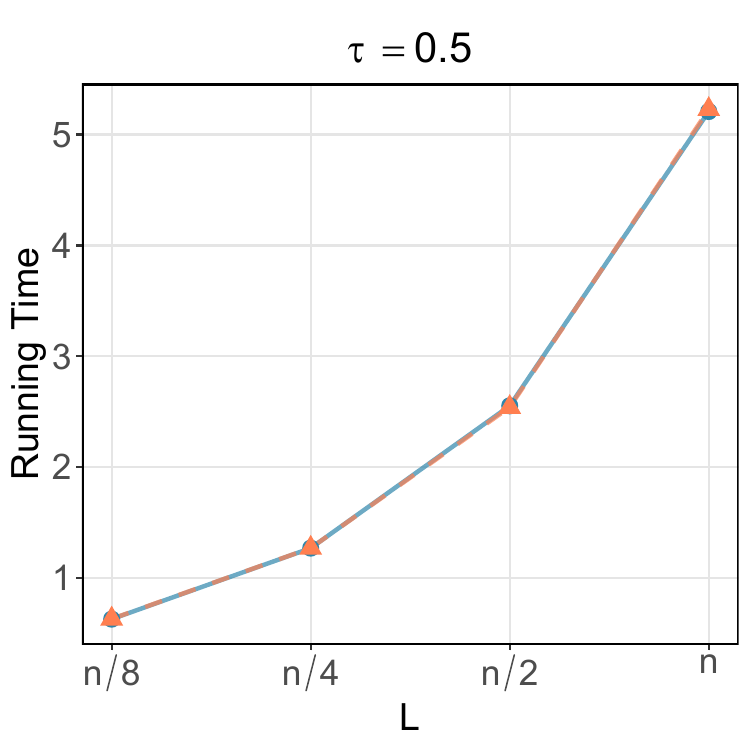}\hfill
\includegraphics[width=.242\linewidth,height=1.75in,keepaspectratio]{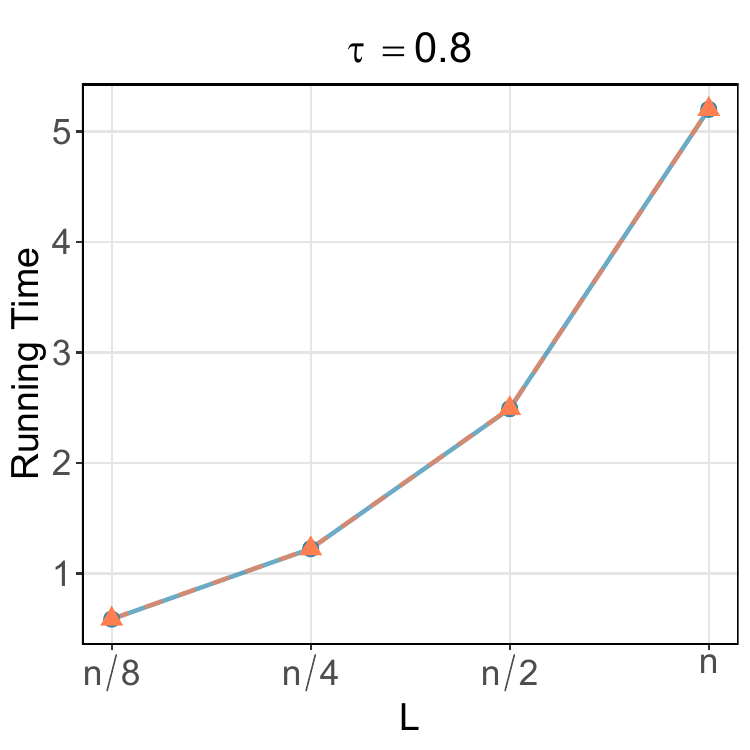}\hfill
\includegraphics[width=.242\linewidth,height=1.75in,keepaspectratio]{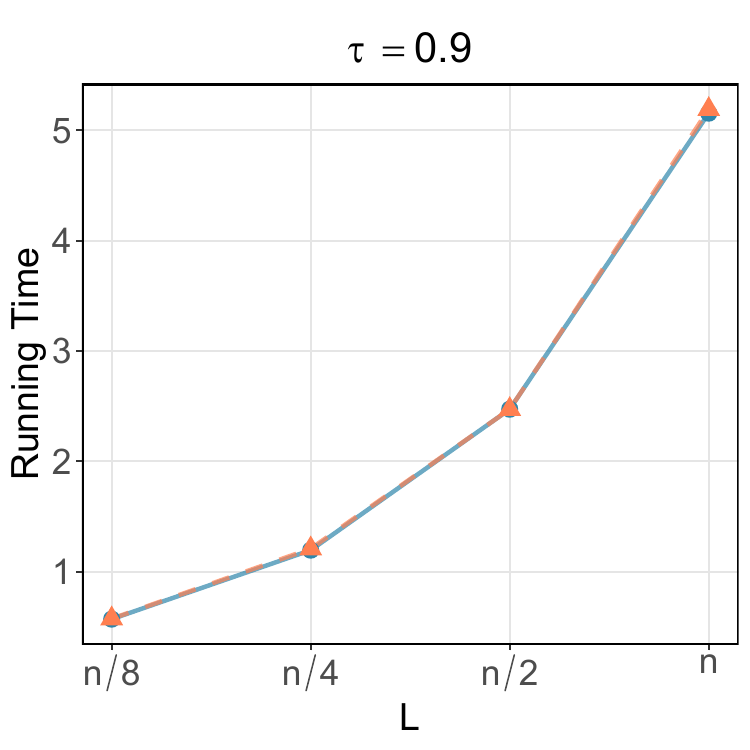}\hfill
\includegraphics[width=.242\linewidth,height=1.75in,keepaspectratio]{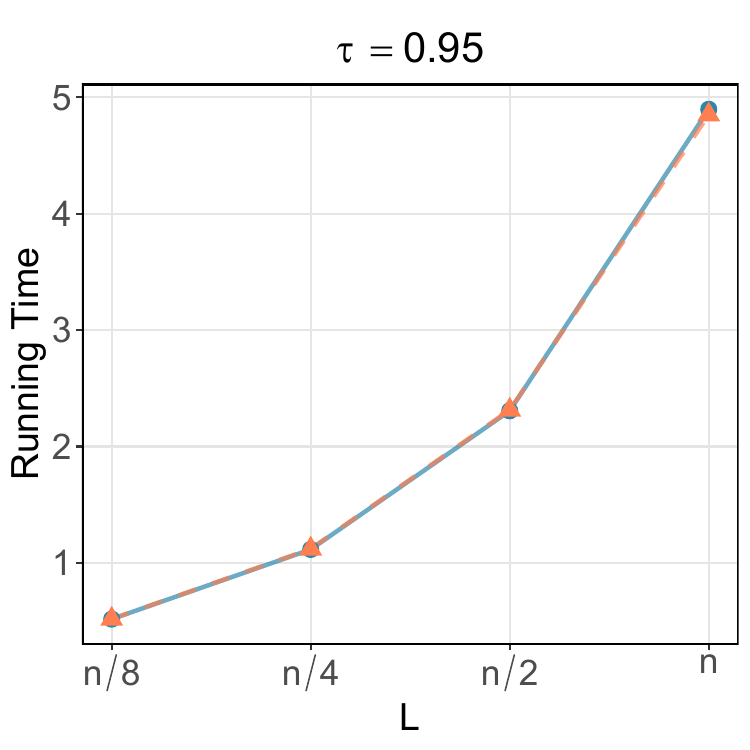}
\caption{Prediction errors and running time (in minutes) of GADMM and CE-GADMM for different numbers of random features, with $L \in \{n,\allowbreak \lfloor n/2\rfloor,\allowbreak \lfloor n/4 \rfloor,\allowbreak \lfloor n/8 \rfloor\}$. }\label{fig:real_rf}
\end{figure}

We first investigate the effect of the number of RFs. Specifically, we consider $L \in \{n,\allowbreak \lfloor n/2\rfloor,\allowbreak \lfloor n/4 \rfloor,\allowbreak \lfloor n/8 \rfloor\}$. Figure~\ref{fig:real_rf} shows that using more RFs generally
improves prediction accuracy, whereas reducing $L$ substantially lowers the
computational cost. Across the four quantile levels,
$L=\lfloor n/2\rfloor$ retains most of the predictive accuracy obtained with
$L=n$ while requiring markedly less computation. Based on this accuracy-efficiency trade-off, we set $L=\lfloor n/2\rfloor$ for subsequent comparisons. The curves for GADMM and CE-GADMM are close throughout, indicating that the communication filter causes little loss of predictive accuracy.

\begin{figure}[!ht]
\centering
\includegraphics[width=.242\linewidth,height=1.75in,keepaspectratio]{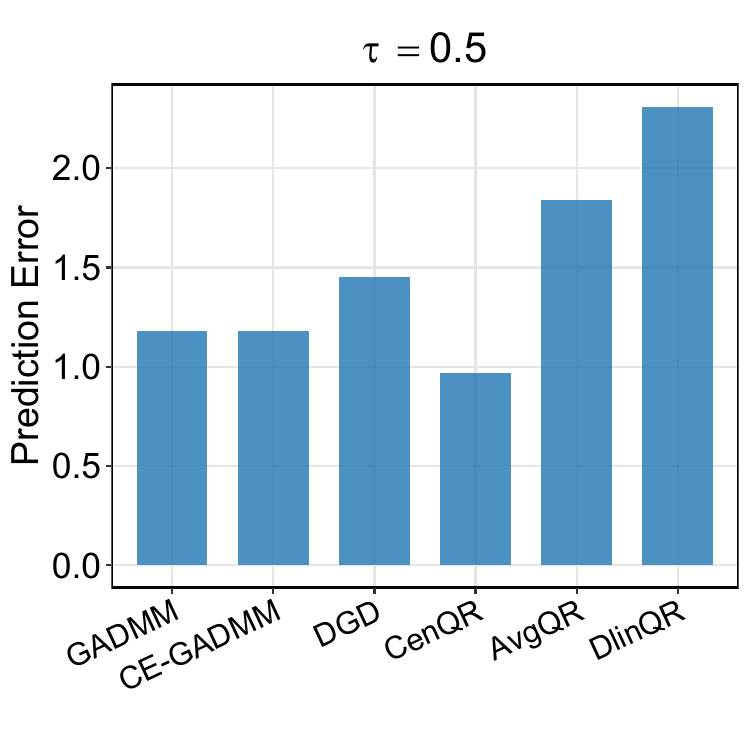}\hfill
\includegraphics[width=.242\linewidth,height=1.75in,keepaspectratio]{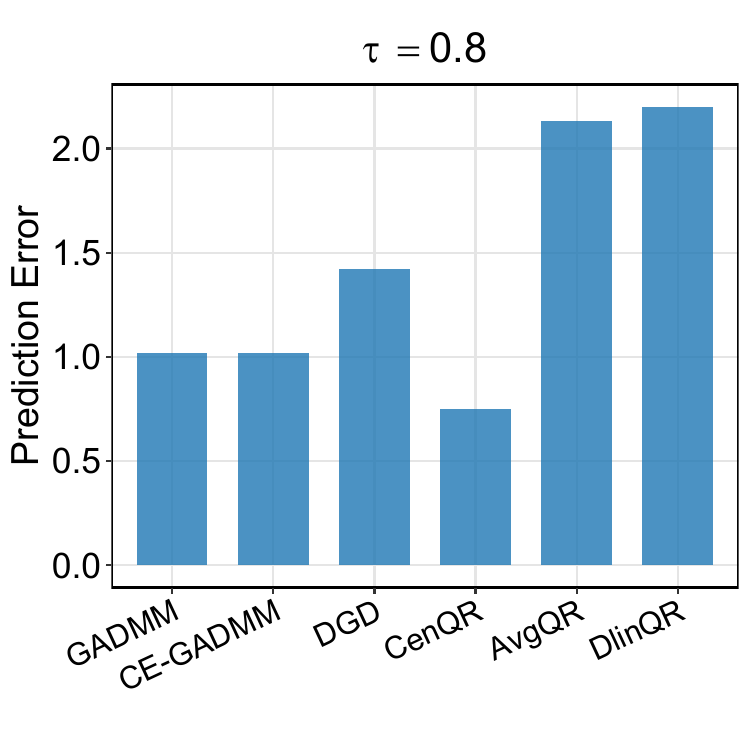}\hfill
\includegraphics[width=.242\linewidth,height=1.75in,keepaspectratio]{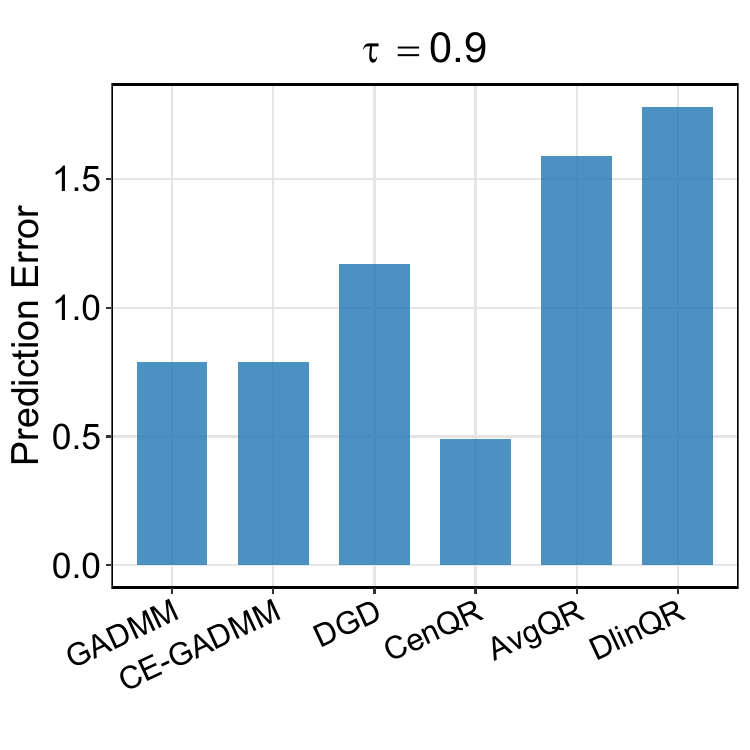}\hfill
\includegraphics[width=.242\linewidth,height=1.75in,keepaspectratio]{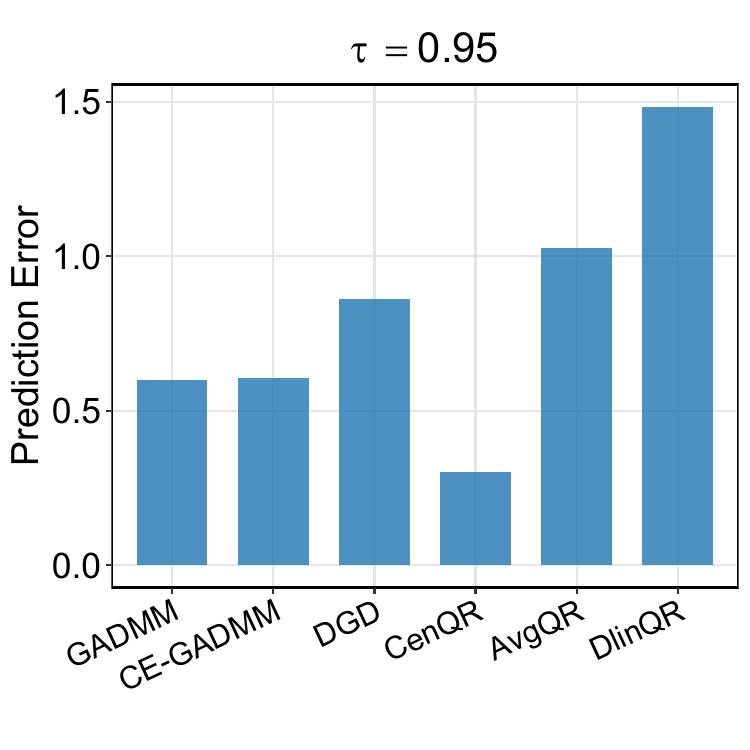}\par\medskip
\includegraphics[width=.46\linewidth]{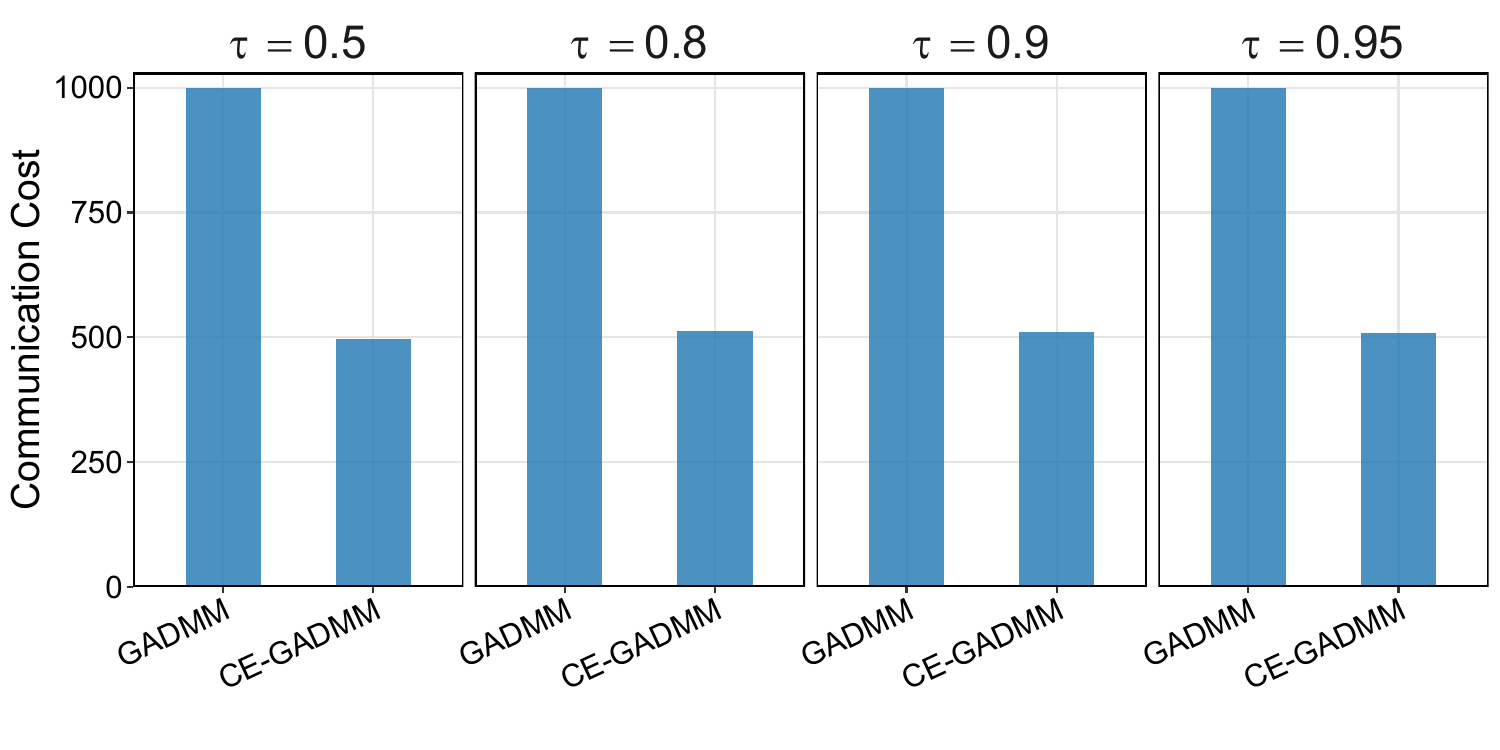}
\caption{Comparison with different methods. }
\label{fig:real_compare}
\end{figure}

We next compare GADMM and CE-GADMM with DGD, CenQR, AvgQR, and the decentralized linear quantile regression method of \citet{wang2018distributed} (denoted by DlinQR). From Figure~\ref{fig:real_compare}, the proposed methods attain predictive performance  comparable to that of the centralized benchmark CenQR, without requiring the exchange of raw observations.
In contrast, DGD and naive averaging of local estimators (AvgQR) provide only limited improvement.
CE-GADMM achieves a favorable communication-accuracy trade-off.
By avoiding transmissions of redundant updates, CE-GADMM substantially reduces communication overhead by approximately 50\% while maintaining nearly identical predictive performance to GADMM.
This property is particularly valuable for large-scale environmental monitoring networks, where frequent communication among geographically distributed stations may be costly or infeasible. DlinQR exhibits the poorest predictive performance suggesting that linear quantile models cannot adequately capture the complex nonlinear relationships between PM$_{2.5}$ and meteorological and pollutant variables, thereby highlighting the importance of flexible nonlinear modelling in this application.

\begin{figure}[!ht]
\centering
\includegraphics[width=.242\linewidth,height=1.75in,keepaspectratio]{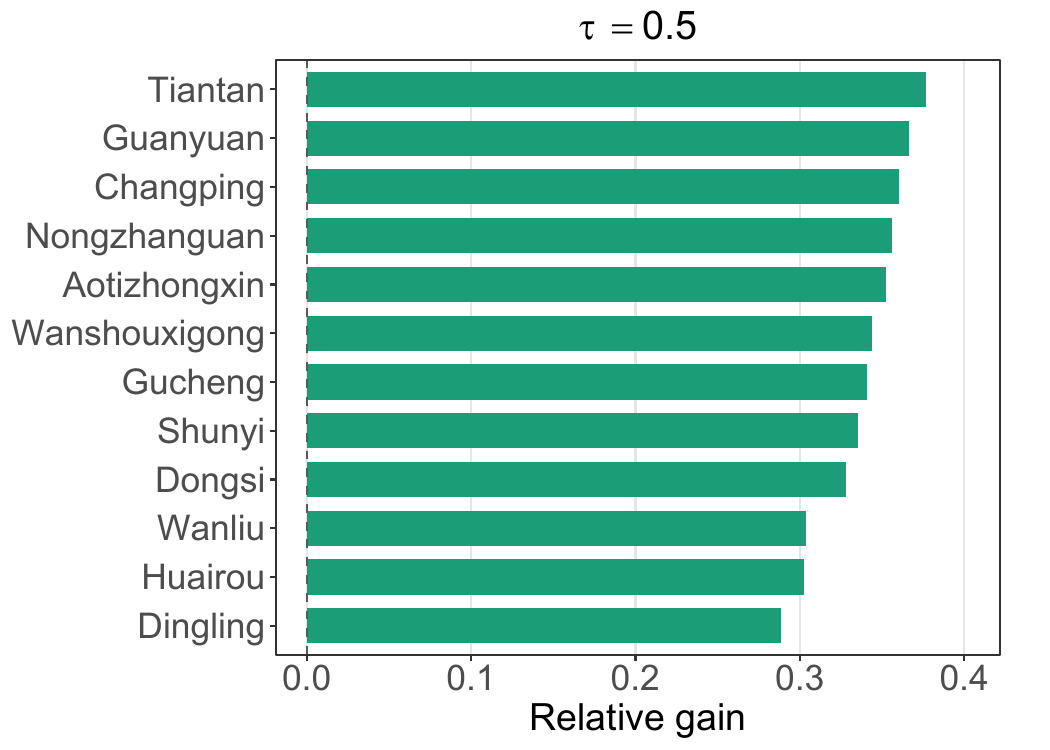}\hfill
\includegraphics[width=.242\linewidth,height=1.75in,keepaspectratio]{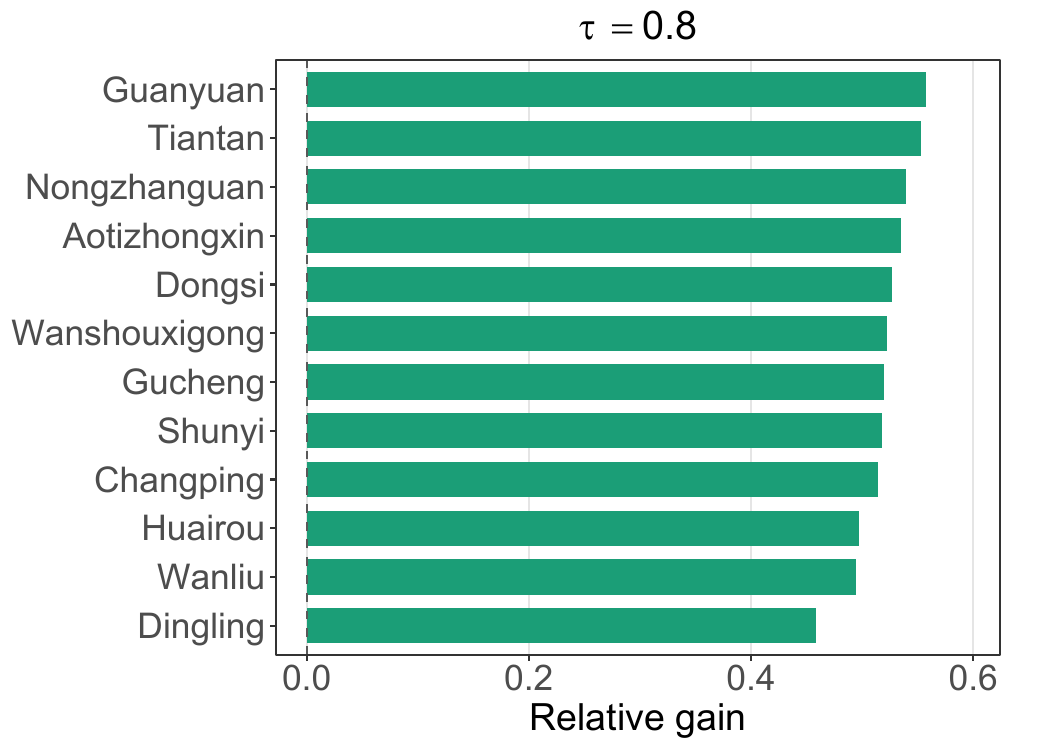}\hfill
\includegraphics[width=.242\linewidth,height=1.75in,keepaspectratio]{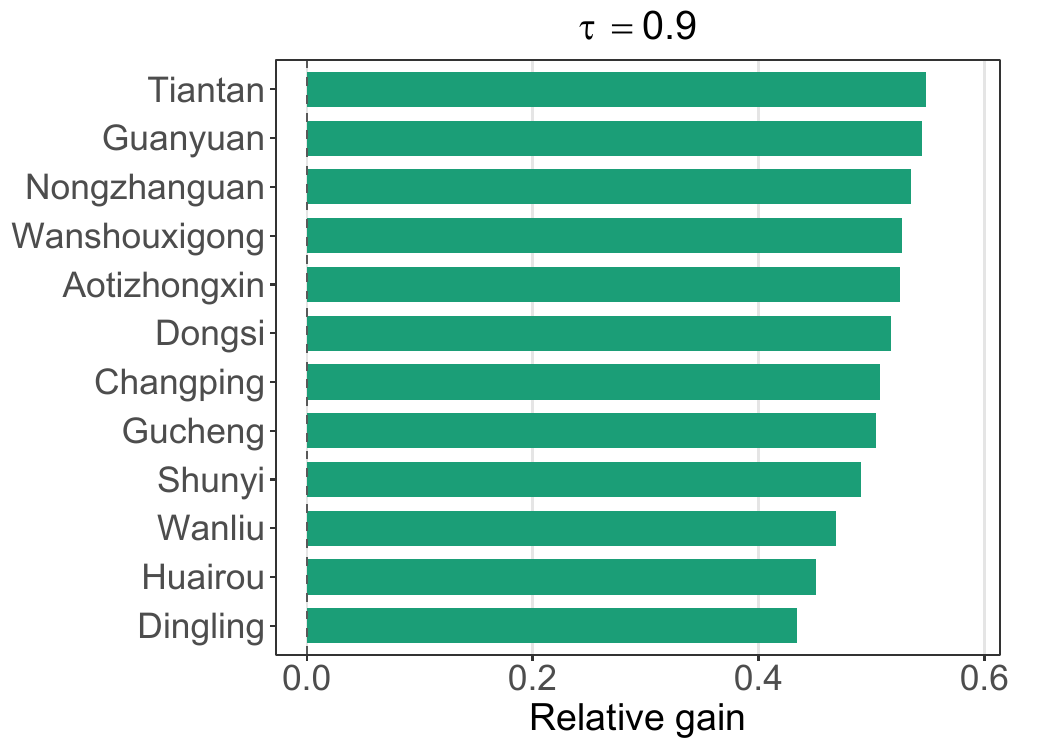}\hfill
\includegraphics[width=.242\linewidth,height=1.75in,keepaspectratio]{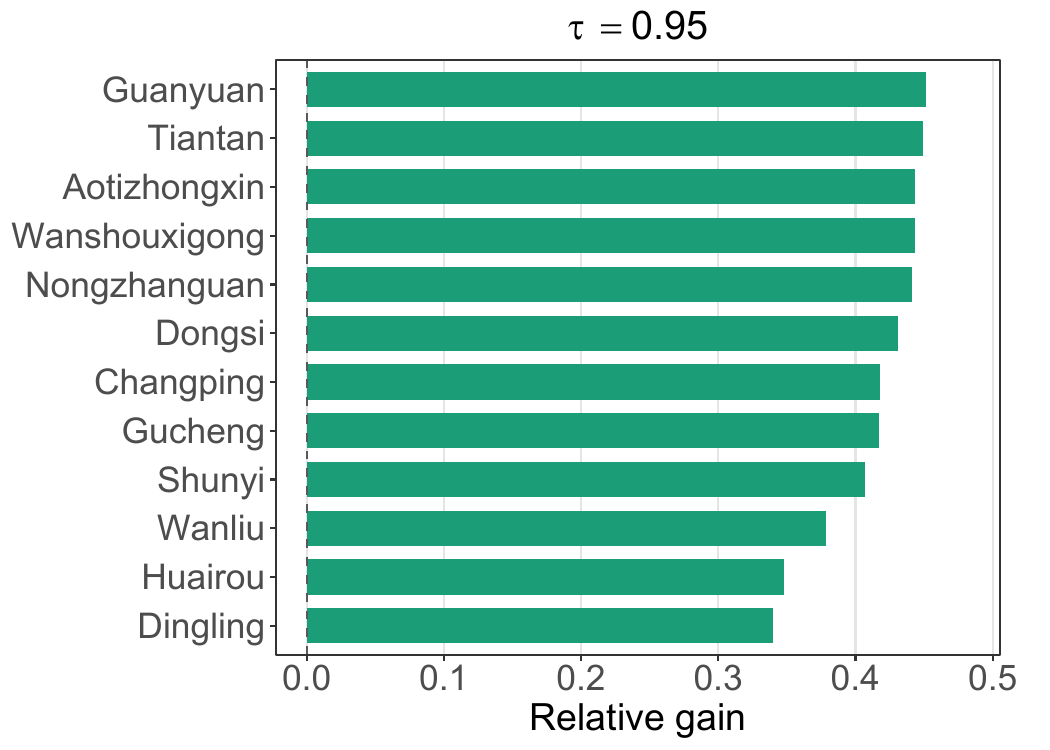}
\caption{Station-specific relative prediction gains of GADMM over LocQR. }\label{fig:local_benefit}
\end{figure}

To examine whether all stations benefit from decentralized collaboration, we further evaluate station-specific prediction performances in Figure~\ref{fig:local_benefit}.  For station $j$, define the relative prediction gain of GADMM over local estimation method LocQR (LocQR denotes the local QR carried out by a single agent) as
\begin{equation*}
\operatorname{RelGain}_{j}
= \frac{\operatorname{PreErr}_{j}^{\mathrm{LocQR}}
-\operatorname{PreErr}_{j}^{\mathrm{GADMM}}}{\operatorname{PreErr}_{j}^{\mathrm{LocQR}}}\,,
\end{equation*}
where a positive value indicates that decentralized collaboration reduces the prediction error relative to fitting each station independently. Comparing these relative gains across stations provides a direct measure of where information sharing is most beneficial.  Figure~\ref{fig:local_benefit} shows that the relative gains are at least around 30\% for all 12 monitoring stations across different quantile levels, demonstrating that every station benefits substantially  from decentralized collaboration. This universal improvement confirms that information sharing effectively alleviates the limited sample size available at individual stations while exploiting the spatial dependence among neighbouring monitoring sites. The magnitude of the relative gains, however, varies across stations. Urban stations, including Tiantan and Guanyuan consistently achieve the largest improvements, whereas suburban stations such as Dingling and Huairou benefit substantially less. This pattern is consistent with the spatial distribution of the monitoring network. Monitoring stations are much more densely deployed in the urban area, resulting in shorter distances between neighbouring stations and consequently stronger spatial dependence in pollution dynamics. Such dense local connectivity enables GADMM to borrow more information from neighbouring stations. In contrast, suburban stations are geographically farther apart and are more likely to experience heterogeneous emission sources and meteorological environments, reducing the amount of transferable information available through decentralized collaboration.

\begin{figure}[!ht]
\centering
\includegraphics[width=.242\linewidth,height=1.75in,keepaspectratio]{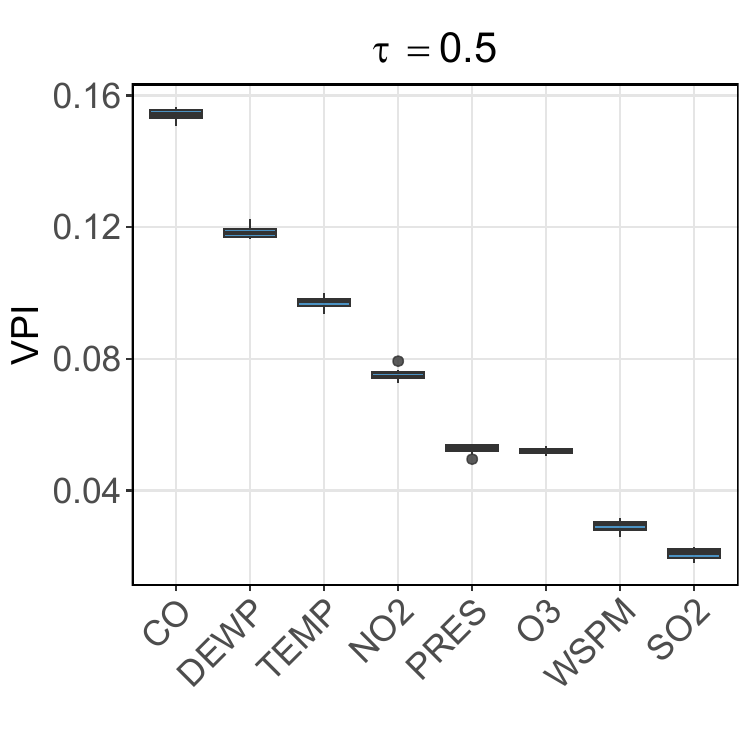}\hfill
\includegraphics[width=.242\linewidth,height=1.75in,keepaspectratio]{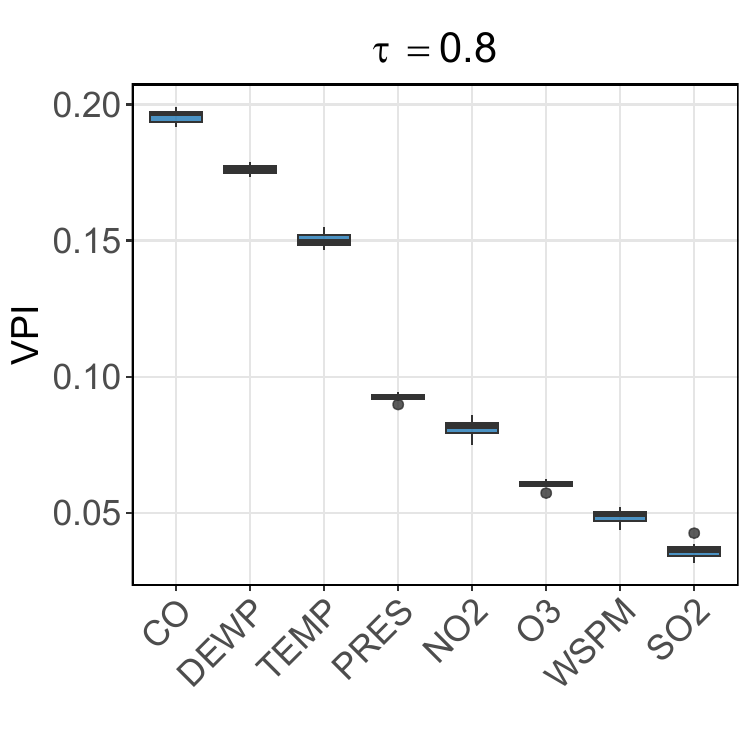}\hfill
\includegraphics[width=.242\linewidth,height=1.75in,keepaspectratio]{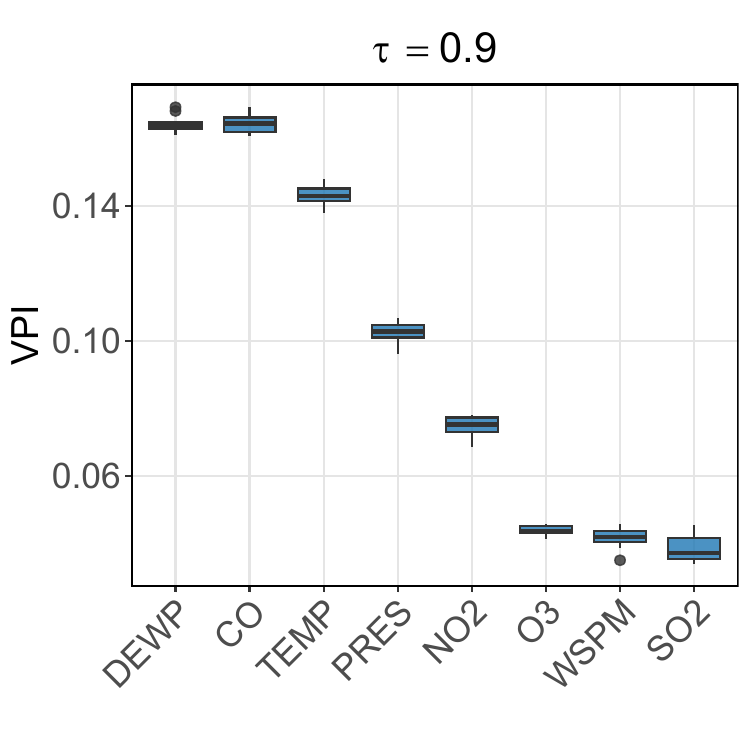}\hfill
\includegraphics[width=.242\linewidth,height=1.75in,keepaspectratio]{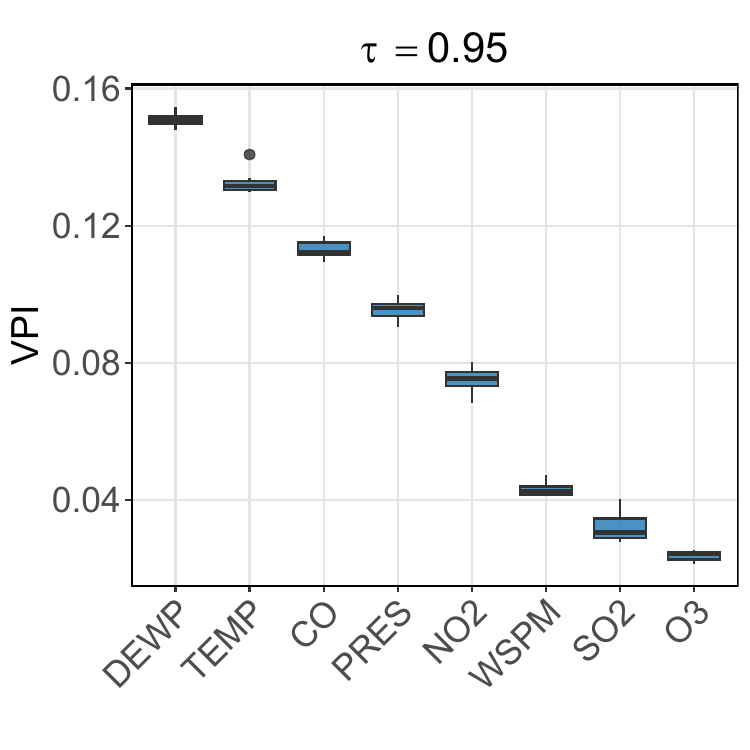}
\caption{Variable permutation importance of GADMM at different quantile levels.}\label{fig:VPI}
\end{figure}

Finally, we assess the contribution of each predictor using variable
permutation importance (VPI). For a given predictor, its VPI is defined as the relative increase in prediction loss after randomly permuting that predictor while leaving all other predictors unchanged  \citep{breiman2001random}. Therefore, a larger VPI value indicates a greater contribution to predicting PM$_{2.5}$. Figure~\ref{fig:VPI} presents boxplots of the VPI values. The results reveal that among all predictors, \texttt{CO}, \texttt{DEWP}, and \texttt{TEMP} consistently appear as the three most influential variables. At $\tau=0.5$ and $0.8$, \texttt{CO} exhibits the largest VPI, indicating that combustion-related
emissions are the dominant factor in explaining typical and moderately
high PM$_{2.5}$ concentrations. This observation is consistent with the
fact that \texttt{CO} is closely associated with primary particulate
emissions caused by human activities such as transport-related emissions and
heating.
As the quantile level increases, however, meteorological variables become increasingly important relative to direct emission indicators. At $\tau=0.9$, \texttt{DEWP} surpasses \texttt{CO}, while at $\tau=0.95$, \texttt{DEWP} and \texttt{TEMP} become the two most important predictors.
High \texttt{DEWP} indicate high humidity, which can promote secondary aerosol formation. Meanwhile, \texttt{TEMP} affects atmospheric stability, chemical reaction rates, and vertical mixing processes, thereby influencing the accumulation and dispersion of particulate matter.  These findings highlight the advantage of quantile analysis over mean-based approaches by revealing pollution mechanisms that vary across the conditional distribution of PM$_{2.5}$. They also suggest that emission-control measures are particularly effective for ordinary pollution levels, whereas mitigating severe pollution episodes requires greater consideration of meteorological conditions.

\bibliographystyle{plainnat}
\bibliography{Bibliography-MM-MC}

\end{document}

%% file: self-define.tex
\def\trans{^{\rm T}}

\def\wt{\widetilde}
\def\wh{\widehat}

\def\lan{\langle}
\def\ran{\rangle}

\newcommand{\nn}{\nonumber}

\definecolor{myblue}{HTML}{1F77B4}
\definecolor{myorange}{HTML}{FF7F0E}

\def\bse{\begin{eqnarray*}}
\def\ese{\end{eqnarray*}}
\def\be{\begin{eqnarray}}
\def\ee{\end{eqnarray}}

\def\vzero{{\bf 0}}
\newcommand{\veins}{{\bf 1}}

\newcommand{\vr}{{\bf r}}

\newcommand{\vu}{{\bf u}}
\newcommand{\vv}{{\bf v}}
\newcommand{\vk}{{\boldsymbol{k}}}
\newcommand{\vw}{{\bf w}}
\newcommand{\vx}{{\bf x}}

\newcommand{\vz}{{\bf z}}

\newcommand{\vA}{{\bf A}}
\newcommand{\vB}{{\bf B}}

\newcommand{\vI}{{\bf I}}

\newcommand{\vM}{{\bf M}}

\newcommand{\vP}{{\bf P}}

\newcommand{\vW}{{\bf W}}

\newcommand{\vphi}{\bm{\phi}}
\newcommand{\vomega}{\bm{\omega}}

\newcommand{\vOmega}{\bm{\Omega}}

\newcommand{\valpha}{\bm{\alpha}}
\newcommand{\vbeta}{\bm{\beta}}

\newcommand{\vtheta}{\bm{\theta}}

\newcommand{\vchi}{\bm{\chi}}

\def\calA{{\mathcal A}}

\def\calD{{\mathcal D}}

\def\calH{{\mathcal H}}
\def\calI{{\mathcal I}}

\def\calK{{\mathcal K}}
\def\calL{{\mathcal L}}

\def\calN{{\mathcal N}}

\def\calS{{\mathcal S}}
\def\calT{{\mathcal T}}

\def\calV{{\mathcal V}}

\def\calX{{\mathcal X}}